\documentclass[%
 aip,
 amsmath,amssymb,
 reprint,%
]{revtex4-1}
\pdfoutput=1

\usepackage{dcolumn}
\usepackage{bm}

\usepackage{graphicx}
\usepackage[space]{grffile}
\usepackage{latexsym}
\usepackage{textcomp}
\usepackage{longtable}
\usepackage{tabulary}
\usepackage{booktabs,array,multirow}
\usepackage{amsfonts,amsmath,amssymb}
\usepackage{natbib}
\usepackage{url}
\usepackage{hyperref}
\hypersetup{colorlinks=true,pdfborder={0 0 0},citecolor=blue}
\usepackage{etoolbox}
\newif\iflatexml\latexmlfalse

\AtBeginDocument{\DeclareGraphicsExtensions{.pdf,.PDF,.eps,.EPS,.png,.PNG,.tif,.TIF,.jpg,.JPG,.jpeg,.JPEG}}

\usepackage[utf8]{inputenc}
\usepackage[greek,english]{babel}
\usepackage[mediumspace,mediumqspace,Grey,squaren]{SIunits}

\begin{document}

\title{A computational perspective of the role of Thalamus in cognition}

\author{Nima Dehghani}
\email{nima.dehghani@mit.edu}
\affiliation{Department of Physics, Massachusetts Institute of Technology}
\affiliation{Center for Brains, Minds and Machines (CBMM), Massachusetts Institute of Technology}
\author{Ralf D. Wimmer}
\affiliation{Department of Brain and Cognitive Sciences, Massachusetts Institute of Technology}

\date{\today}

\begin{abstract}
Thalamus has traditionally been considered as only a relay source of cortical inputs, with hierarchically organized cortical circuits serially transforming thalamic signals to cognitively-relevant representations. Given the absence of local excitatory connections within the thalamus, the notion of thalamic `relay' seemed like a reasonable description over the last several decades. Recent advances in experimental approaches and theory provide a broader perspective on the role of the thalamus in cognitively-relevant cortical computations, and suggest that only a subset of thalamic circuit motifs fit the relay description. Here, we discuss this perspective and highlight the potential role for the thalamus -- and specifically mediodorsal (MD) nucleus -- in dynamic selection of cortical representations through a combination of intrinsic thalamic computations and output signals that change cortical network functional parameters. We suggest that through the contextual modulation of cortical computation, thalamus and cortex jointly optimize the information/cost trade-off in an emergent fashion. We emphasize that coordinated experimental and theoretical efforts will provide a path to understanding the role of the thalamus in cognition, along with an understanding to augment cognitive capacity in health and disease. 
\end{abstract}%

\keywords{Thalamo-cortical system, Recurrent Neural Network, Reservoir Computing, Multi-objective Optimization, Cognitive Computing, Artificial Intelligence}

\maketitle



\section*{Cortico-centric view of perceptual and cognitive processing}


Until recently, cognition {[}in mammalian, bird and reptilian nervous system{]} has been viewed as a cortico-centric process, with thalamus considered to only play the mere role of a relay system. This classic view, much driven by the visual hierarchical model of the mammalian cortex \citep*{Felleman_1991}, puts thalamus at the beginning of a feedforward hierarchy. The transmission of information from thalamus to early sensory cortex (V1 in the visual system for example), and the gradual increasing complex representations from V2 to MT/IT and eventually prefrontal cortex (PFC), constitute the core of the perceptual representation under the hierarchical model. A recent comparative study of biologically-plausible Convolutional Neural Networks (CNN) and the visual ventral stream, emphasizes on the feature enrichment through the network of hierarchically interconnected computational modules (layers in the neural network and areas in the ventral stream) \citep{Yamins_2014}.

\begin{widetext}

\begin{figure}[h!]
\begin{center}
\includegraphics[width=0.7\columnwidth]{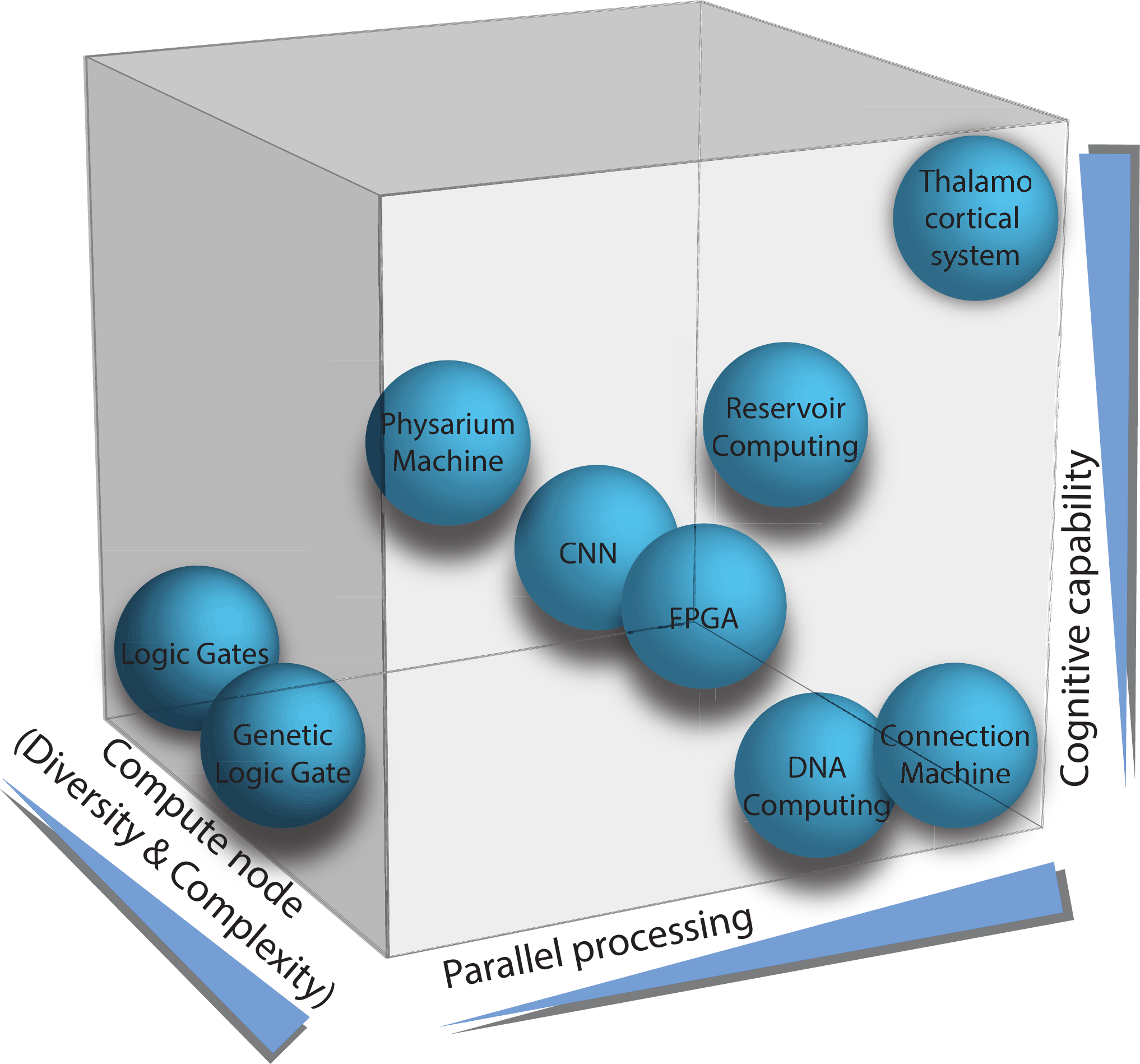}
\caption{{\textbf{Cognitive computing morphospace:} Morphospace of a few example
biological and synthetic computing engines in a multidimensional layout.
The thalamo-cortical system standout as a unique system with high
cognitive capacity, massive parallel processing and extreme diversity of
the computing nodes. Other computing systems occupy less desirable
domains of this morphospace. \emph{Logic Gates}: NAND, NOR;
\emph{Genetic Logic Gate}: Synthetic biology adaptation of logic gate;
\emph{FPGA}: field-programmable gate array (configurable integrated
circuit); \emph{CNN}: Convolutional Neural Network; \emph{Physarium
Machine}: programmable amorphous biological computer experimentally
implemented in slime mould. \emph{Reservoir Computing}: A reservoir of
recurrent neurons that dynamically change their activity to nonlinearly
map the input to a new space; \emph{DNA computing}: A computing paradigm
where many different DNA molecules are used to perform large number of
logical computations in parallel; \emph{Connection Machine}: the first
commercial supercomputer designed for problems of Artificial
Intelligence (AI) using hardware enabled parallel processing.  For a multi-dimensional representation of global/local computation, serial/parallel processing and complex/simple computation, see \citep{Sipper1999}. For an idealized landscape of computation involving the degree of relevance of space, agent (cell) diversity and  distributed computing see \citep{Sole2013}.
{\label{morphospace}}%
}}
\end{center}
\end{figure}
\end{widetext}

The strictly static feedforward model has since morphed to a dynamic hierarchical model due to the discoveries of the role of feedback from higher cortical areas to lower cortical areas \citep{Heeger2017}. These dynamic hierarchies are even considered to be favorable for recursive optimizations, where the overall optimization can be achieved by breaking the problem into smaller ones to find the optimum for each of these smaller problems. Such recursive optimization does not need to be confined to just one cortical area and each of such distributed optimizations may even be solved differently. \citep*{Marblestone2016}. This view of cortical computation is also paralleled with the growing use of recurrent neural networks (RNN) that can capture the dynamics of single neurons or neural population in a variety of tasks. As such, RNNs can mimic (a) context-dependent prefrontal response \citep{Mante2013} or (b) can reproduce the temporal scaling of neural responses in medial frontal cortex and caudate
nucleus \citep{Wang_2017}. Although it is clear that higher level cortical feedback reaches all the way down to thalamus (Wimmer et al, other refs), the main attribute of perception/cognition remains cortico-centric under the umbrella of dynamic hierarchical models or RNN embodiment of cortical cognitive functions. Since the computation that is carried by the system should match the computing elements at the appropriate scale \cite{Dehghani2018} a mismatch between these presumed computational systems and the underlying circuitry becomes vividly apparent. First, associative cortex (and not just sensory cortex) receives thalamic input. Second, certain thalamic territories receive their primary (driving) input from the cortex, rather than sensory periphery, some of which are likely to be highly convergent on a single thalamic cell level, suggesting at least a cortical modulation of sensory input. Third, thalamic projections can be broad and diffuse, suggesting a modulatory, rather than a relay function. These points are indicative that thalamus may play a central part in cognitive processes. But what does including thalamus add that could not be achieved in cortical loops? One advantage that the thalamus could bring into the cortical equation is flexibility. For example, the same sensory cue may have different meanings according to a particular context a subject is in. A recent study \cite{Rikhye2018} shows that thalamus may be well suited to reorganize functional connectivity in frontal cortices in response to such contextual changes, allowing for a more flexible switching of rule to action mapping. We suggest that the unique cognitive capability of the thalamo-cortical system is tightly bound to parallel processing and contextual modulation that are enabled by the diversity of computing nodes (including thalamic and cortical structures) and complexity of the computing architecture (Fig.~{\ref{morphospace}}). We will start with a brief overview of the thalamic architecture, followed by experimental evidence and a computational perspective of the thalamic role in contextual cognitive computation. 


\section*{Thalamic architecture: anatomical and functional features}

\begin{figure}[h!]
\begin{center}
\includegraphics[width=1\columnwidth]{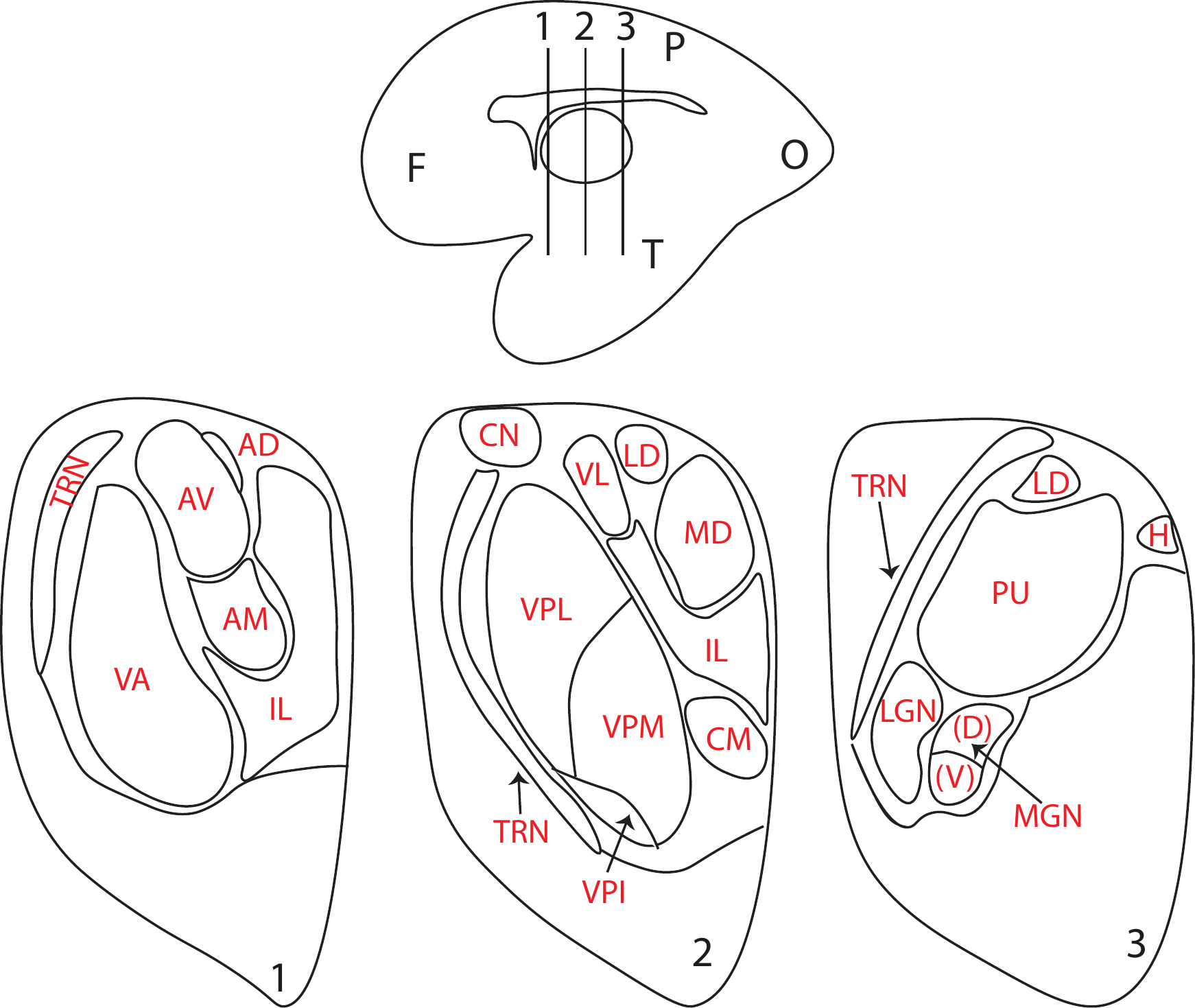}
\caption{{\textbf{Schematic layout of thalamic nuclei:} Three cross sections of
monkey thalamus. \emph{AD}: anterodorsal nucleus; \emph{AM}:
anteromedial nucleus; \emph{AV}: anteroventral nucleus; \emph{CM}:
centromedian nucleus; \emph{CN}: caudate nucleus; \emph{H}: habenular
nucleus; \emph{IL}: intralaminar nuclei;\emph{LD}: lateral dorsal
nucleus; \emph{LGN}: lateral geniculate nucleus;\emph{MD}: mediodorsal
nucleus; \emph{MGN(D)}: medial geniculate nucleus
(dorsal); \emph{MGN(V)}: medial geniculate nucleus (ventral); \emph{PU}:
pulvinar; \emph{TRN}: thalamic reticular nucleus (not a relay
nucleus); \emph{VA}: ventral anterior nucleus; \emph{VL}: ventral
lateral nucleus; \emph{VPI}: ventral posterior nucleus
(inferior); \emph{VPL}: ventral posterior nucleus (lateral); \emph{VPM}:
ventral posterior nucleus medial. Redrawn from~\protect\citep*{Sherman_2013}.
{\label{thalamus}}%
}}
\end{center}
\end{figure}

Traditionally, thalamic nuclei (see Fig.~{\ref{thalamus}}) are defined as collections of neurons that are segregated by gross features such as white matter tracts and various types of tissue staining procedures \citep{Jones1981}. This gross anatomical classification has been equated with a functional one, where individual thalamic nuclei giving rise to a set of defined functions \cite{Jones1981,Jones1985}. More recent fine anatomical studies challenge this notion, showing that within individual nuclei, single cell input/output connectivity patterns are quite variable.

\begin{widetext}

\begin{figure}[h!]
\begin{center}
\includegraphics[width=0.6\columnwidth]{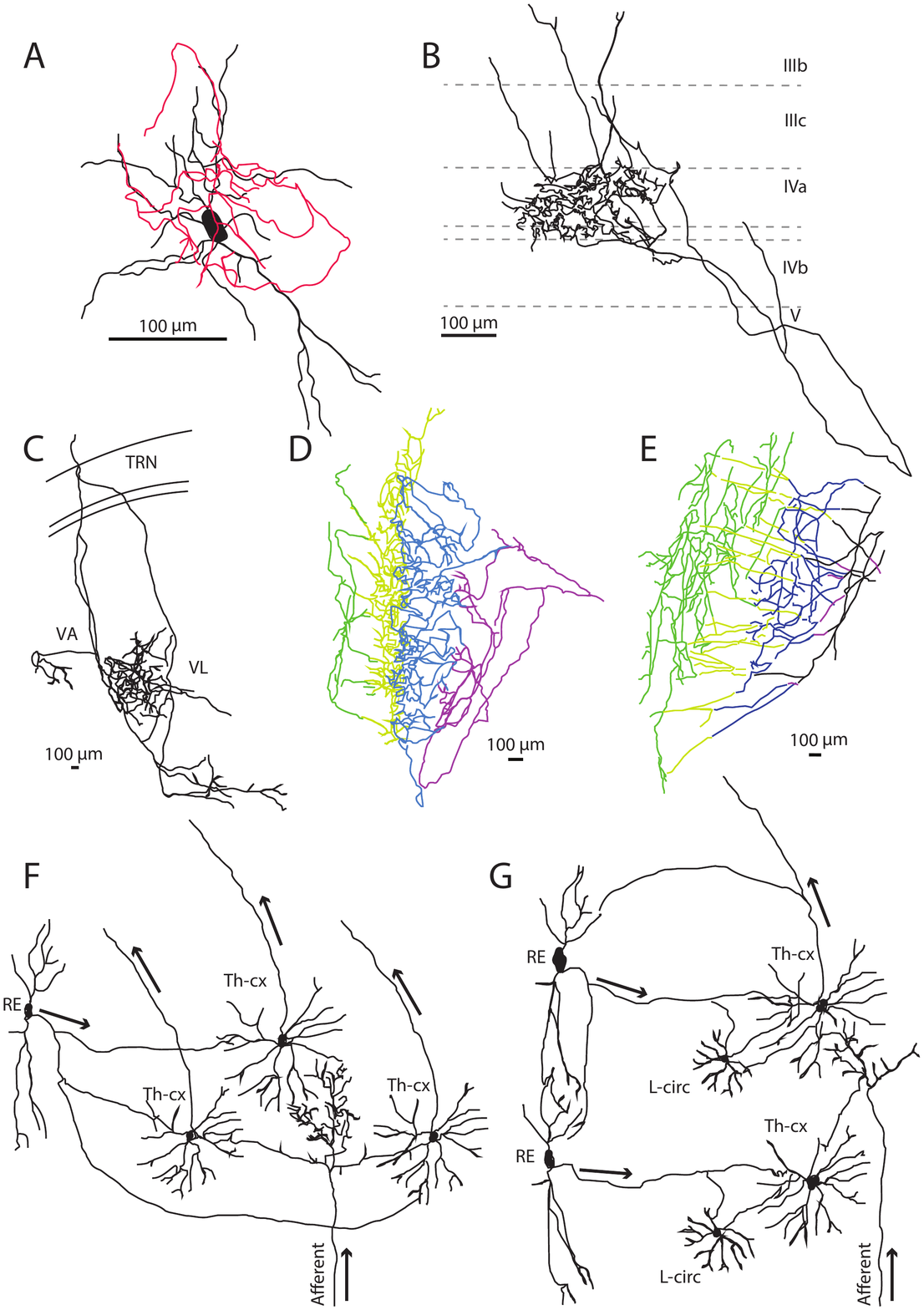}
\caption{{\textbf{Diversity and complexity of thalamo-cortical
architectures:} \emph{(A)} Comparative size of a single RGC (retinal
ganglion cell) terminal (red) and an LGN neuron (black); Redrawn
from \citep{FitzGibbon2015}. \emph{(B)} Complete terminal arbor of a single
LGN neuron projecting to V1; Redrawn from \citep*{Raczkowski1990}. \emph{(C)}
Projection of a TC (thalamo-cortical) neuron to cat motor cortex (with
only 23 terminals in TRN versus 1632 terminals in VA/VL); Redrawn
from \citep{Kakei2001,Ohno2012} \emph{(D,E)} Axonal arborization of a single MD
neuron \emph{(E)} and a single POm neuron \emph{(E)}. Principal target
layers are: layer I (green), layer II-IV (green-yellow), layer V (blue),
layer VI (purple); Redrawn from \citep{Kuramoto2016}. Note the comparative
size of panels \emph{A-E} (scale bar at \unit{100}{\micro\meter}). \emph{(F,G)} Synaptic
network of thalamo-cortical (Th-cx), Reticular thalamic cell (RE) and
local circuit (L-circ) thalamic interneuron in rodents \emph{(F)} and
feline/primates \emph{(G)}. Note that rodents do not have L-circ.
Afferent axon excites Th-cx, which in return sends the signal to
cortex. RE inhibitory effect on Th-cx cells varies depending on the
excitatory drive to each Th-cx cell (\emph{F}: compare the two neurons
on the right versus the one on the left). Axonal collaterals of an RE
cell could inhibit another RE cell (\emph{G}: top RE inhibits the bottom
RE), which releases the activity of L-circ leading to inhibition of
weakly excited Th-cx (bottom) adjacent to the active Th-cx (top).
Panels \emph{F} and \emph{G} are redrawn from \citep*{Steriade2007} based
on experiments from \citep{Steriade1984,Steriade1986}.
{\label{cells}}%
}}
\end{center}
\end{figure}
\end{widetext}

Thalamus lacks lateral excitatory connections and rather receives inputs from other subcortical structures and/or cortex. In fact, a major feature of forebrain expansion across evolution is the invasion of the thalamus by cortical inputs \citep{Grant_2012,Rouiller_2000}. Most (90-95\%) afferents to the relay nuclei are not from the sensory organs \citep{Jones1985,Sherman_2013}. Recent anatomical studies have shown a great diversity of cortical input type, strength and inferred degree of convergence, even within individual thalamic nuclei \citep{Clasca2012} (see Fig.~{\ref{cells}} for an example view of cell and network architecture diversity).

Excitatory inputs, mostly, arrive as feedbacks from layer 6 of the cortex as well as from brainstem reticular formation. In addition to the diversity of excitatory inputs, thalamic circuits receive a diverse set of inhibitory inputs (note that local GABAergic thalamic interneurons are mostly absent in non-LGN relay nuclei). The two major systems of inhibitory control are the thalamic reticular nucleus (TRN), a shell of inhibitory nucleus surrounding thalamic excitatory nuclei, and the extra-thalamic inhibitory system; a group of inhibitory projects across the fore-, mid- and hindbrain (see \citep*{Halassa_2016} for a review on thalamic inhibition). Perhaps a major differentiating feature of these two systems (TRN and ETI) is temporal precision. One of the key characteristics of thalamus is lack of direct local loops. Only a very small group of inhibitory neurons with local connections exists in thalamus \citep*{Steriade_1988}. A mechanistic consequence of this
architecture is the differential control of thalamic response gain and selectivity (Fig.~{\ref{cells}} F and G), with the TRN controlling the first as observed in sensory systems \citep{Pinault_2004}, and ETI controlling the latter as observed in motor systems \citep*{Urbain_2007}. For example, basal ganglia control of thalamic responses, a form of ETI control, would be implemented through thalamic disinhibition, which is not only dependent on ETI input, but also a special type of thalamic conductance that enables high frequency `bursting' upon release from inhibition \citep{Goldberg_2013,Deniau_1985}. 

Overall, the variety of thalamic inputs (both excitatory and inhibitory), combined with intrinsic thalamic features such as excitability and morphology will determine the type of intrinsic computations that thalamus performs. This view appears to be consistent with recent observations of confidence encoding in both sensory \citep{Komura_2013} and motor systems \citep{Ma_2014,Jazayeri_2015}. 


Thalamic relay nuclei mostly project to the cortical middle layers in a topographic fashion. However, the majority of thalamic structures also project more diffusely to the cortical superficial layers, such as mediodorsal (MD), posteriodmedial complex (POm) and pulvinar for example (see Fig.~{\ref{cells}} for an example of thalamic cell and circuit diversity). These diffuse projections seem poorly suited to relay information in a precise manner. Rather, they might have a modulatory role of cortical function. Further, a great degree of diversity can be observed at the level of thalamic axonal terminals within the cortex. While the idea of a thalamic relay was consolidated by observing that the main LGN neurons thought to be associated with form vision (M and P pathways) exhibit spatially compact cortical terminals, recent anatomical studies of individual neurons across the thalamus show a variety of terminal sizes and degree of spatial spread and intricate computational architecture (Fig.~{\ref{cells}}). This complexity of the architecture and diversity of the computing nodes are among the key factors that set apart the thalamo-cortical system from other conventional and unconventional computing engines (Fig.~{\ref{morphospace}}). Part of the complication in understanding how these anatomical types give rise to different functions is their potential for contacting different sets of excitatory and inhibitory cortical neurons. 

Specifically, among thalamic nuclei, mediodorsal thalamus (MD) seems to have a connectivity pattern that is distinctively different from the classic sensory nuclei. Cortico-thalamic projections to MD originate from both layers V and VI of PFC \citep{Goldman1985,Giguere1988,Mitchell2015}. But, in contrast to relay nuclei, cortical input to MD terminates both extraglomerularly and within the synaptic glomeruli, suggesting that cortex plays a different role in MD activity (in comparison to LGN for example) \citep{Schwartz1991}. Additionally, optogenetic and \textit{in vitro} electrophysiology techniques have revealed that MD not only projects to the layer I but also has additional terminations in layer III \citep{Cruikshank2012}. MD projections to PFC synapse with both excitatory and inhibitory cells \citep{Cruikshank2012} where the triggered triggered feedforward inhibition could play a variety of roles from regulating dendritic action potentials \citep{Kim1995} to imposing a narrow temporal window within which the excitation can reach the target \citep{Cruikshank2007}.  These elaborate input and output connectivities of the thalamocortical architecture point to the non-unitary computational role of thalamus. Among thalamic nuclei, specifically MD (and likely other MD-like nuclei) has the architecture necessary to be involved in modulatory computational roles rather than the well-known relay functions. In the next sections, we provide the experimental evidence and algorithmic designs pointing to this modulatory role. This idea of the thalamus controlling cortical state parameters is highlighted in Figs.~{\ref{experiment}},~{\ref{computation}} and the next section.

\section*{Many facets of thalamic computation}

It is commonly thought that processes like attention, decision making and working memory are implemented through distributed computations across multiple cortical regions \citep{Corbetta1998,Mesulam1990,Scott2017}. However, it is unclear how these computations are coordinated to drive relevant behavioral outputs. From an anatomical standpoint, the thalamus is strategically positioned to perform this function, but relatively little is known about its broad functional engagement in cognition. The thalamic cellular composition and network structure constrain how cortex receives and processes information. The thalamus is the major input to the cortex and interactions between the two structures are critical for sensation, action and cognition \citep{Jones1998,Sherman2016,Nakajima2017}. Despite recent studies showing that the mammalian thalamus contains several circuit motifs, each with specific input/output characteristics, the thalamus is traditionally viewed as a relay to or between cortical regions \citep*{Sherman_2013}.  That the active role of thalamus in cognition is beyond relay stems from a) the fact that sensory input to thalamus is much limited in comparison to input from other structures, such as cortex and basal ganglia  ,  b) the experimental evidence showing number of nuclei modulating cortical neural processing according to behavioral context and c) lesions of certain nuclei such as pulvinar and MD cause severe attention and memory deficits \citep{Saalmann2015}. Below we will discuss how the distinctive anatomical architecture and computational role of pulvinar and MD differ from relay nuclei such as LGN.

It is worth mentioning that this view of bona fide thalamic computations is quite distinct from the one in which thalamic responses reflect their inputs, with only linear changes in response size. This property of reflecting an input (with only slight modification of amplitude) was initially observed in the lateral geniculate nucleus (LGN), which receives inputs from the retina. LGN responses to specific sensory inputs (their receptive fields, RF) are very similar to those in the retina itself, arguing that there is little intrinsic computation happening in the LGN itself outside of gain control. Success in early vision studies \citep{Hubel1959,Hubel1962} might have inadvertently given rise to the LGN relay function being generalized across the thalamus. The strictly feedforward thalamic role in cognition requires reconsideration \citep*{Halassa2017}; only a few thalamic territories receive peripheral sensory inputs and project cortically in a localized
manner, as the LGN does\citep{Jones1981,Raczkowski1990,Kakei2001,FitzGibbon2015,Sherman2016}.

The largest thalamic structures in mammals, the MD and pulvinar contain
many neurons that receive convergent cortical inputs and project
diffusely across multiple cortical layers and
regions \citep{Clasca2012,Rovo2012}. For example, the primate pulvinar has both
territories that receive topographical, non-convergent inputs from the
striate cortex \citep*{Rovo2012} and others that receive convergent
inputs from non-striate visual cortical (and frontal)
areas \citep*{Mathers_1972}. This same thalamic nucleus also receives
inputs from the superior colliculus \citep{Partlow_1977}, a subcortical
region receiving retinal inputs. This suggests that the pulvinar
contains multiple input `motifs' solely based on the diversity of
excitatory input. Such input diversity is not limited to the pulvinar,
but is seen within many thalamic nuclei across the mammalian
forebrain \citep{Bickford_2016}. Local inactivation of pulvinar neurons
results in reduced neural activity in primary visual
cortex \citep{Purushothaman2012} suggesting a feedforward role. Recent studies, however, indicate that pulvinar may have additional roles. For example, pulvinar inactivation was shown to increase low-frequency cortical oscillations \citep{Zhou2016}. Given that such activity is often associated with inattention and sleep, this study suggested that pulvinar may keep cortical regions in an activated state that would allow responsiveness to top-down input from other areas to modulate ongoing activity according to attentional demands. A different study showed that during perceptual decision making pulvinar
neurons encode choice confidence, rather than stimulus
category \citep{Komura_2013}. Together, these recent findings strongly argue for more pulvinar functions beyond relaying information. 

In the case of MD, direct sensory input is limited \citep{Mitchell2015}
and the diffuse, distributed projections to cortex \citep{Kuramoto2016}
are poorly suited for information relay; this input/output connectivity
suggests different functions. Recent studies \citep{Bolkan_2017,Schmitt_2017, Rikhye2018} have
begun to shed light on the type of computation that MD performs.
Taking advantage of the genetic accessibility of the mouse for neuronal manipulations, these studies have revealed that MD coordinates task-relevant activity in the prefrontal
cortex (PFC) in a manner analogous to a contextual signal regulating
distinct attractor states within a computing reservoir. Specifically, in
a task where animals had to keep a rule in mind over a brief delay
period (Fig.~{\ref{experiment}}A), PFC neurons show
population-level persistent activity following the rule presentation, a
sensory cue that instructs the animal to direct its attention to either
vision or audition (Fig.~{\ref{experiment}}B,C). MD neurons
show responses that are devoid of categorical selectivity
(Fig.~{\ref{experiment}}D), yet are critical for selective
PFC activity; optogenetic MD inhibition diminishes this activity, while
MD activation augments it. The conclusion is that MD inputs enhance
synaptic connectivity among PFC neurons or may adjust the activity of
PFC neurons through selective filtering of the thalamic inputs. In other
words, delay-period MD activity maintains rule-selective PFC
representations by augmenting local excitatory
recurrence \citep{Schmitt_2017}.

In a related study, a delayed nonmatching-to-sample T-maze working
memory task \citep{Bolkan_2017}, it was shown that MD amplification and
maintenance of higher PFC activity indicated correct performance during
the subsequent choice phase of the task. Interestingly, MD-dependent
increased PFC activity was much more pronounced during the later (in
delay) rather than earlier part of the task. These findings indicate
that PFC might have to recursively pull in MD to sustain cortical
representations as working memory weakens with time. Together these
studies indicate that PFC cognitive computation can not be dissociated
from MD activity. Further evidence for the critical role of the MD-PFC
interaction for cognition is the disrupted fronto-thalamic anatomical
and functional connectivity seen in neurodevelopmental
disorders \citep{Parnaudeau_2015,Mitelman_2005,Marenco_2011,Nair_2013,Woodward_2017}. 

\subsubsection*{Can MD select cortical subnetworks based on contextual modulation?}

\begin{widetext}

\begin{figure}[h!]
\begin{center}
\includegraphics[width=0.6\columnwidth]{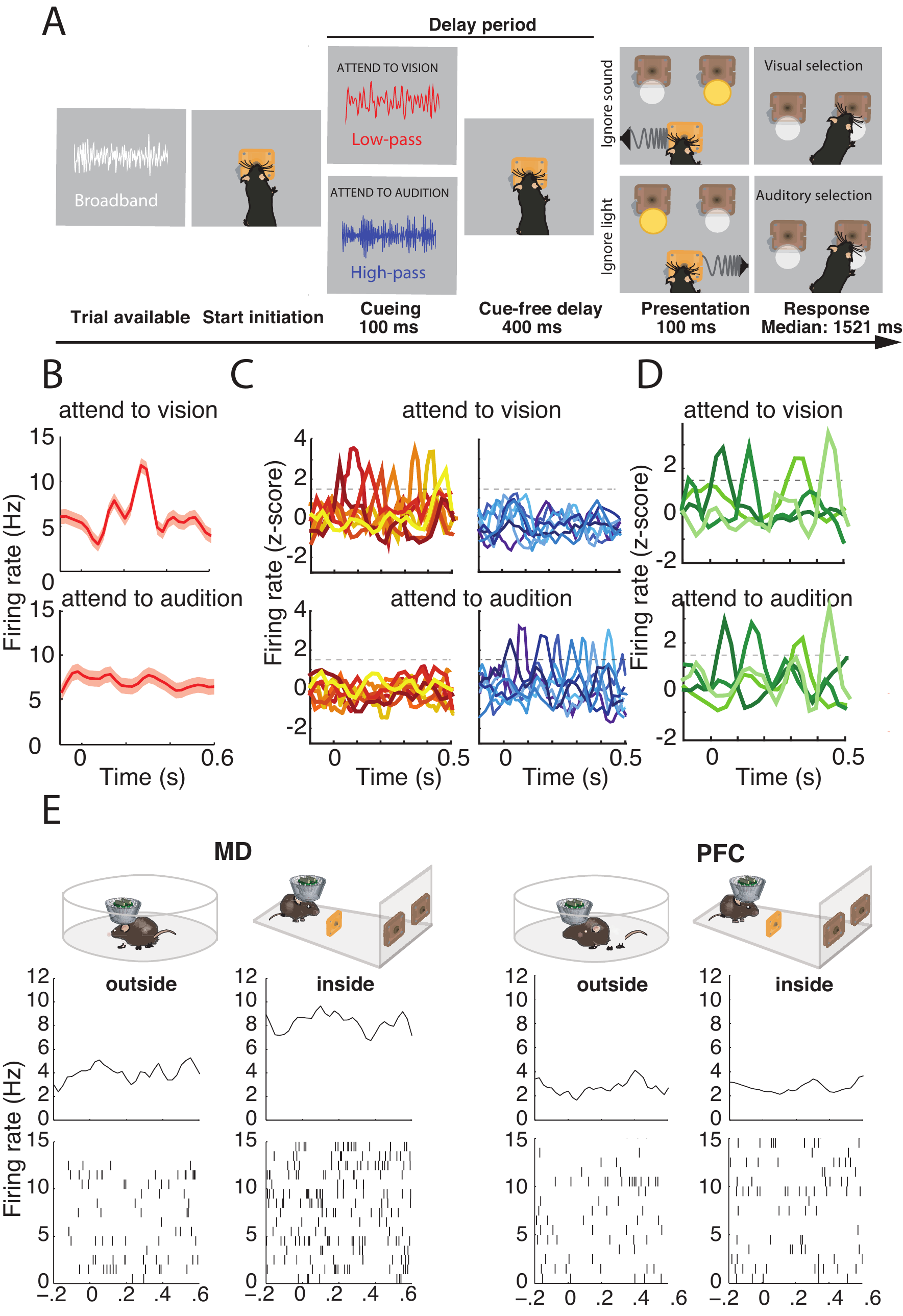}
\caption{{\textbf{MD-PFC interactions during sustained rule
representations.} \emph{(A)} attentional control task design. \emph{(B)}
Example peri-stimulus time histogram (PSTH) for a neuron tuned to attend
to vision rule signaled through low pass noise cue \emph{(C)} Examples
showing that rule-specificity is maintained across distinct PFC
rule-tuned populations. \emph{(D)} PSTHs of four MD neurons showing
consistent lack of rule specificity. \emph{(E)} Example rasters and
PSTHs of an MD and PFC neuron when the animal is engaged in the task and
outside of the behavioral arena. In contrast to PFC, MD neurons show the
contextual difference in a change in firing rate. Figure is redrawn
from \citep{Schmitt2017}.
{\label{experiment}}%
}}
\end{center}
\end{figure}
\end{widetext}

Why would a recurrent network (PFC) computation depend on its
interaction with a non-recurrent (MD) non-relay network? What
computational advantage such system would have? Using a chemogenetic
approach, a recent study suggested that information flow in the MD-PFC
network can be unidirectional. While both inactivating PFC-to-thalamus
and MD-to-cortex pathways impaired recognition of a change in reward
value in rats performing a decision making task, only the inactivation
of MD-to-cortex pathway had an impact on the behavioral response to a
change in action-reward relationship \citep{Alcaraz_2018}. Given that a
sensory stimulus may require a different action depending on the context
in which it occurs, the ability to flexibly re-route the active PFC
subnetwork to a different output may be crucial. In an architecture like
the PFC-MD network, where MD can modulate PFC functional connectivity,
MD might well be suited to re-route the ongoing activity in a context
dependent manner. In fact, in the mouse cognitive task described above
(Fig.~{\ref{experiment}}A), a subset of MD neurons showed
substantial spike rate modulation during task engagement compared to
when the animals is in its home in cage (see
Fig.~{\ref{experiment}}E) \cite{Schmitt_2017}. In contrast,
PFC neurons show very little difference in spike rates when the animal
gets engaged in the task. This suggests that perhaps different subsets
of MD neurons are capable of encoding task `contexts', which has been shown experimentally \citep{Rikhye2018}. Subsequently,
each given subset could unlock a distinct cortical association; this hypothesis is now experimentally verified \citep{Rikhye2018}. These MD
subsets have to be able to shift the cortical states dynamically while
maintaining the selectivity based on the subset of cortical connections
they target. This idea would also fit with the paradigm shift indicating
that thalamic neurons exert dynamical control over information relay to
cortex \citep{Basso_2005,Parnaudeau_2017}.

Overall, the anatomical and neurophysiological data show that the
thalamic structure and cortico-thalamic network circuitry, and the
interplay between thalamus and cortex, shape the frame within which
thalamus plays the dual role of relay and modulator. Under this
framework, different thalamic nuclei carry out multitude of functions
including but not limited to information relay. A suggestion of this
comparative computational role of LGN, pulvinar and MD is depicted in
Fig~{\ref{viewpoint}}. The importance of (non-relay) thalamic nuclei's regulatory influence on cortical function is also reflected by the disorders that emerge due to thalamic dysfunctions. Specifically, lesions to pulvinar and MD lead to severe attention and memory deficits \cite{Saalmann2011, Baxter2013}. The disruption of MD-PFC communication is the likely cause of these cognitive impairment. As mentioned earlier, the back and forth interaction between MD and PFC is necessary during the task acquisition period and is reflected by an increase in (beta frequency) MD-PFC synchrony \citep{Parnaudeau2013}. In addition, MD also regulates the neural synchrony of PFC neurons \citep{Saalmann2014}. Decreasing MD spiking activity (by hyperpolarizing MD neurons) leads to disrupted MD-PFC synchrony and impaired performance in the delayed non-match to sample task \citep{Parnaudeau2013}. Moreover, schizophrenics show both a reduced beta and gamma frequency deficit \citep{Uhlhaas2010}, significant reduction of MD volume \citep{Popken2000,Alelupaz2008} and total number of MD neurons \citep{Popken2000}. While it remains unclear whether the MD loss of neuron is primary or secondary to PFC pathology \citep{Popken2000}, it is evident that MD is regulating PFC plasticity, cognitive flexibility \citep{Baxter2013} and contextual processing \citep{Schmitt_2017,Rikhye2018}.

\section*{Is thalamus a read-write medium for cortical parallel processing?}

The connectivity pattern of relay and non-relay point to the dichotomy of algorithmic constrains that are imposed by these specific thalamic structures. There are stark contrasts between sensory (LGN-like) versus non-sensory thalamic nuclei thalamocortical and cortico-thalamic connectivity profiles. Anatomical tracing and radiographic studies have shown that while relay nuclei have preserved topographic focal projections the to middle cortical layer, major thalamic structures (MD, Pulvinar, POm) have a more diffuse projection to the superficial layers of cortex \citep{Krettek1977, Giguere1988, Mitchell2015}. Among non-relay nuclei, MD shows interesting characteristics, projecting not only to the layer I but also to the outer banks of layer III \citep{Cruikshank2012}. Only a small fraction of MD to PFC projections end in middle cortical layers and more than 90$\%$ have modulatory and projectdissuely to superficial layers of PFC \citep{Viaene2011a, Viaene2011b, Zikopoulos2007}. In addition, cortico-thalamic axons to non-relay nuclei show a dual role for cortical influence on thalamus \citep{Giguere1988,Mitchell2015}.  For example, not only layer VI/V PFC project huge number of axons directly to MD but also send collaterals to the reticular thalamic nucleus and indirectly influence MD activity \citep{Giguere1988,Schwartz1991,Sherman_2002}. These massive reciprocal connectivity is metabolically very costly for the brain. If brain were to operate as a simple pattern matching system, a far less costly feed-forward network (similar to the structure suggested by \citep{Yamins_2014}) would have been more economical. The complex connectivity of non-relay thalamus and cortex point to computations that are beyond pattern matching. This view also matches what we know of cortical computation, namely that it involves multiple sources of expertise (each decoding certain aspect of the incoming stimuli), along with the fact that these expertise modules operate in a highly parallel mode. To coordinate these parallel yet convergent cortical processing modules, there is a need for a system that is commonly visible to each module. In addition, collectively, these modules should keep track of the changes in the stimuli in the environment and integrate the information in time. This contextual processing would require small modifications of processed information in individual modules. Or for handling the local constraints, it may need repetition of certain algorithms until the satisfactory results are reached. The connectivity profile that we discussed provides the platform to run such computations.

If the brain were to function as a simple pattern matching system
without wiring and metabolic constrains, evolution would just expand the
size and depth of the network to the point that it could potentially
memorize a large number of possible patterns. Possibly, evolution would
have achieved this approximation of arbitrary patterns by evolving a
deep network. This would be a desirable solution since any system can be
defined as a polynomial Hamiltonians of low order, which can be
accurately approximated by neural networks \citep{Lin_2017}. But
cognition is much more than template matching and classification
achieved by a neural network. The limits of template matching methods in
dealing with (rotation, translation and scale) invariance in object
recognition quickly became known to neuroscientists and in early works
on computer vision. One of the early pioneers of AI, Oliver Selfridge,
proposed Pandemonium architecture to overcome this
issue \citep{Selfridge1959}. Selfridge envisioned serially connected
distinct demons (an image demon, followed by a set of parallel feature
demons, followed by a set of parallel cognitive demons and eventually a
decision demon), that independently perceive parts of the input
before reaching a consensus together through a mixture of serial
culmination of evidence from parallel processing. This simple
feedforward computational pattern recognition model is (in some ways) a
predecessor to modern day connectionist feedforward neural networks,
much like what we discussed earlier in the text. However, despite its
simplicity, Pandemonium was a leap forward in understanding that the
intensity of (independent parallel) activity along with a need to a
summation inference are the keys to move from simple template matching
to a system that has a concept about the processed input. A later
extension of this idea was proposed by Allen Newell as the Blackboard
model: \emph{``Metaphorically, we can think of a set of workers, all
looking at the same blackboard: each is able to read everything that is
on it and to judge when he has something worthwhile to add to it. This
conception is just that of Selfridge's Pandemonium: a set of demons
independently looking at the total situation and shrieking in proportion
to what they see that fits their natures''} \citep{Newell1962}.
Blackboard AI systems, adapted based on this model, have a common
knowledge base (blackboard) that is iteratively updated (written to and
read from) by a group of knowledge sources (specialist modules), and a
control shell (organizing the updates by knowledge
sources) \citep{Nii1986}.

Interestingly, this computational metaphor can also be extended to the
interaction between thalamus and cortex, though thalamic blackboard is
not a passive one as in the blackboard systems \citep{Harth1987,Mumford1991,Mumford1992}.
Although, initially the active blackboard was used as an analogy for LGN
computation, the nature of MD connectivity and its communication with
cortex seem much more suitable to the type of computations that is
enabled by an active blackboard. Starting with an input, thalamus as the
common blackboard visible to processing (cortical) modules, initially
presents the problem (input) for parallel processing by modules. Here by
module, we refer to a group of cortical neurons that form a functional
assembly which may or may not be clustered together (in a column for
example). By iteratively reading (via thalamo-cortical projections) from
and writing (via cortico-thalamic projections) to this active
blackboard, expert pattern recognition modules, gradually refine their
initial guess based on their internal processing and the updates of the
common knowledge. This process continues until the problem is solved
(Fig.~{\ref{computation}}). 

This iterative communication between non-relay thalamus and cortex suggests that cortico-thalamic projections return the results of computations (that was carried in parallel cortical modules) back to the thalamus. The integration of these revisions in the next thalamic output happens via the synaptic input and dendritic arbors of the non-relay thalamic neurons. One of the major differences in synaptic organization of MD from that of the sensory nuclei is that cortical axons target MD neurons both extraglomerularly and within the synaptic glomeruli \citep{Schwartz1991}. In sensory nuclei, these within-glomeruli synaptic sites are particularly designated for receiving major ascending sensory afferents \citep{Spacek1974}. In MD, such large terminals may even engulf multiple synaptic contacts \citep{Pelzer2017} and are positioned on proximal dendrites of thalamic neurons \citep{Schwartz1991}. These within-glomuerli multi-synaptic structures exhibit fast kinetics, large postsynaptic currents and strong short-term depression \citep{Pelzer2017}. Interestingly, the short-term depression is combined with fast recovery after repetitive stimulation \citep{Pelzer2017}, enabling generation of synaptic activity patterns that can match the frequency of cortico-thalamic inputs arriving via these large terminals \citep{Steriade1984b}. As a result, these potent synaptic structures provide the platform for PFC inputs to play a much more active role in shaping the thalamic response in relay (MD) nuclei in comparison to the role that cortical feedback to sensory thalamic nuclei may play \citep{Schwartz1991}.

Distinctive biophysical characteristics of non-relay thalamocortical projections play a complementary role in the computational scheme echoing an active blackboard. Interestingly, activating MD does not generate spikes across a population of prefrontal cortical neurons they project to, while activating LGN generates spikes in primary visual cortex \citep{Schmitt_2017, Rikhye2018}. Instead, MD activation results in overall enhancement of inhibitory tone, coupled with enhanced local recurrent connectivity within the PFC. Although thalamocortical feed-forward inhibition is also observed in somatosensory barrel cortex in response to thalamic stimulation \citep{Gabernet2005}, MD-evoked inhibition in PFC exerts a more powerful inhibitory gain control \citep{Delevich2015}. This difference must be rooted in the particular cortical target pattern of MD projections. Specifically, MD directly targets parvalbumin-positive PV (and not somatostatin --SOM) interneurons in layer I and III \citep{Kuroda1998, Kuroda2004,Rotaru2005,Delevich2015,Collins2018}, while (for example) VM only ``weakly'' activates variety of layer I interneurons \citep{Cruikshank2012}. In fact, when SOM interneuros are silenced, MD-evoked feedforward inhibition is enhanced \citep{Delevich2015}. As a result, while VM plays a more important role in excitation/inhibition balance, MD plays the modulator role via varying the integration time window and temporal precision of cortical responses  \citep{Collins2018}. While individual pyramidal neurons harbor a broad response dynamics, the feedforward inhibition regulates the population dynamic via graded recruitment of individual neurons \citep{Khubieh2016}. Increased conductance, noisy voltage fluctuations, and depolarization are the not-necessarily exclusive factors that define how the changes in the background input, i.e. stimulus changes and their contextual relevance, affect the gain \citep{Cardin2008,Prescott2003}. In addition, while sensory thalamocortical synapses (i.e. LGN to visual cortex) onto fast-spiking inhibitory neurons manifest much higher release probability than those onto pyramidal cells, MD projections show similar presynaptic release probability among the two inhibitory and excitatory cortical neurons \citep{Delevich2015}. The co-variation of excitatory and feed-forward inhibitory response sets the control for graded recruitment of pyramidal neurons into population response \citep{Khubieh2016} . This control in itself is evoked by prior cortical excitation of MD, and as a result, the altered activity of PV interneurons in PFC can bias the response towards passive vs flexible contextual processing in a manner that is distinctively different from the observed response of the sensory cortices \citep{Delevich2015}. These mechanisms show us why the chemogenetic inhibition of MD leads to impaired working memory and flexible goal-directed behaviors \citep{Parnaudeau2013,Parnaudeau_2015}. Similarly, schizophrenics show reduced MD-PFC functional coupling \citep{Mitelman_2005} and deficits in prefrontal PV interneurons \citep{Lewis2012}, highlighting the importance of the modulatory effect of MD on PFc function \citep{Delevich2015}.



\section{Computational and metabolic constrains}
\begin{figure}[h!]
\begin{center}
\includegraphics[width=1\columnwidth]{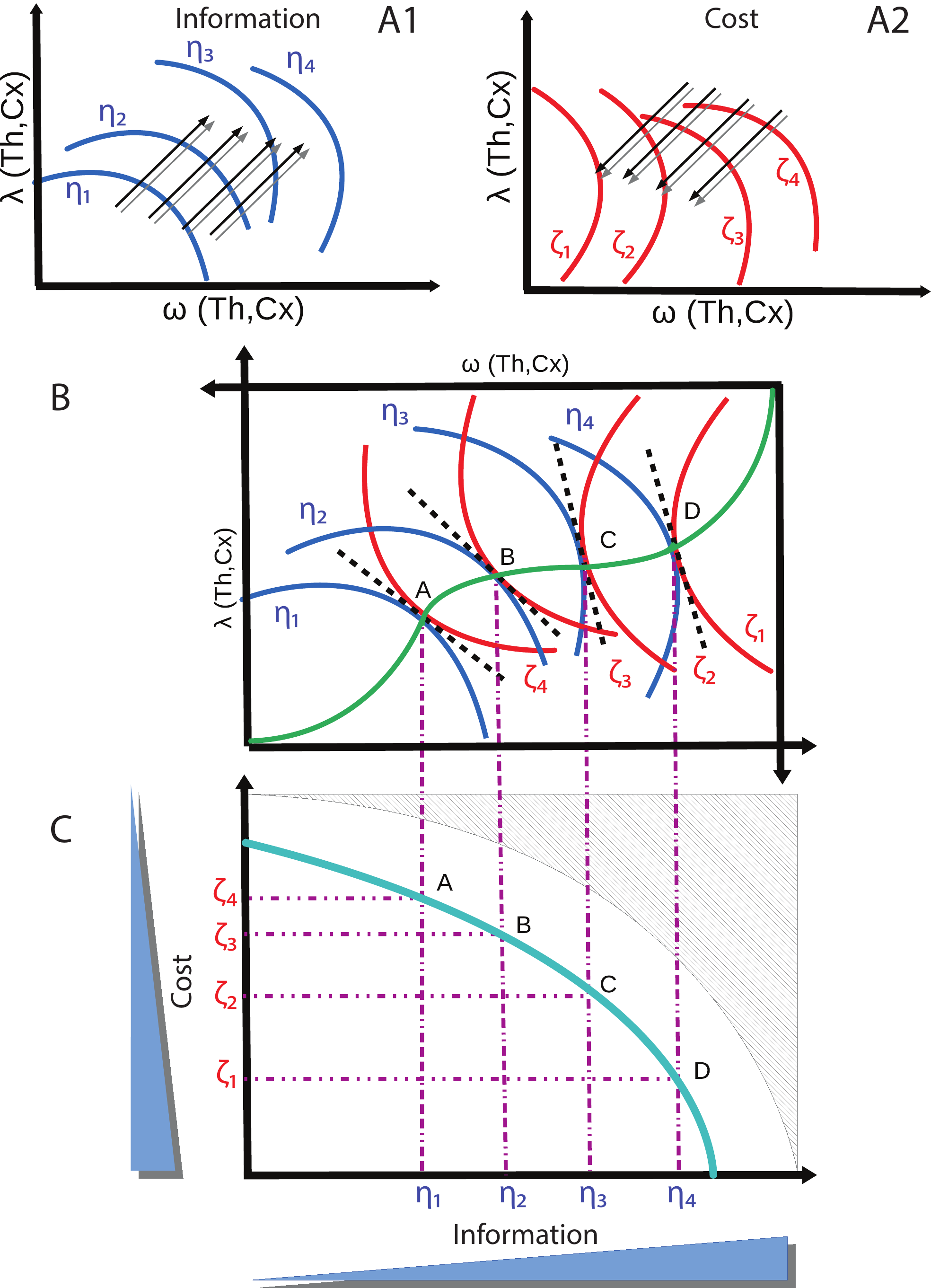}
\caption{{\textbf{Dynamic role of thalamo-cortical system in the information/cost
optimization.} \emph{(A)} Iso-maps of information \emph{(A1)} and cost
\emph{(A2)} in the domain specified
by \(\omega\) and \(\lambda\) (functions of cortical and
thalamic activity). Information across each Iso-quant curve
(\(\eta_1\) for example) is constant and is achieved at a
certain mixture of \(\omega\) and \(\lambda\). Optimal
information can be obtained by moving outward (arrows, \emph{A1}). Cost
optimization can be achieved by moving inward (\emph{A2}). \emph{(B)}
since information and cost are both defined in the domain
of \(\omega\) and \(\lambda\), thalamus and cortex
jointly contribute to information and cost optimization. The points
where the iso-quant curves' tangents are equal (black dashed line),
provide the optimal combination of information/cost (green curve). In
any cycle of cognitive operation, depending on the prior state of the
system (\(\omega\) and \(\lambda\)), the nearest points
on the green curve are the optimal solutions for ending that
cycle. \emph{(C)} mapping of the optima curve to information/cost domain
shows all pareto efficient allocations (cyan curve). The slope of the
parto frontier shows how the system trades cost versus information:
along the pareto curve, efficiency is constant but the exchange between
information and cost is not. All allocations inside of this curve could
be improved as thalamus and cortex interact. The grey zone shows the
biophysically not-permissible allocation of computation and resource.
{\label{theory}}%
}}
\end{center}
\end{figure}

To process the changing stimuli and altering contextual cues, and in order to achieve cognitive flexibility, the thalamocortical system needs to harbor temporal buffering. The mechanisms that we described here point to the ways in which such buffer may takes place, namely: a) changes in the stimuli/context are constantly reprocessed by the cortex and the outputs are rewritten to thalamus, b) thalamus constantly reshapes the cortical population dynamics. MD is changing the mode by which PFC neurons interact with one another, initiating and updating different attractor dynamics underlying distinct cognitive inputs. As a result, the thalamocortical system, collectively and at any instant, keeps an updated description of the stimulus/context over some computational cycles up to the present (Fig.~{\ref{computation}}). Perhaps the upper bound of these computational cycles is tightly bound to the particulars of the cortico-thalamic and thalamocortical connectivity and biophysical constrains. However, since we are dealing with a biological system with finite resources, this back and forth communication needs to have certain characteristics to provide a viable computational solution. First and foremost, the control of interaction and its scheduling has to have a plausible biological component and should bind solutions as time evolves. Second, to avoid turning into an NP-hard (non-deterministic polynomial-time hardness) problem, there must exist a mechanism that stops this iterative computation once an approximation has been reached (Fig.~{\ref{computation}}). Here, we propose a specific solution to the first problem and a plausible one for the later issue. We suggest that phase-dependent contextual modulation serves to deal with the first issue and a multi-objective optimization of efficiency (computational information gain) and economy (computational cost, i.e. metabolic needs and the required time for computation) handles the second issue (Fig.~{\ref{theory}}). In both cases, we suggest that thalamus plays an integral role in conjunction with cortex.\\

\subsection*{Computational constrains and the role of thalamus in phase-dependent contextual modulation}

As mentioned earlier, we know that hierarchical convolutional neural networks (HCNN), which can recapitulate certain properties of static hierarchical forward models, can not capture any processes that need to store prior states \citep*{Yamins2016}. As a result, context-dependent processing can be extremely hard to implement in neural network models \citep{Rigotti2010}. The most widely used ANNs (Feedforward nets , i.e. multilayer perceptrons/Deep Learning algorithms) face fundamental deficiencies: the ubiquitous training algorithms (such as back-propagation), i) have no biophysical plausibility, ii) have high computational cost (number of operations and speed), and iii) require millions of examples for proper adjustment of the connections' weights. These features render feedforward NNs not suitable for temporal information processing. In contrast, recurrent neural networks (RNNs) can universally approximate the state of dynamical systems \citep*{Funahashi1993}, and because of their dynamical memory are well suited for contextual computation. If the higher cortical areas were to show some features of RNN-like networks, as manifested by the dynamical response of single neurons \citep{Mante2013}, then we anticipate that the local computation (interaction between neighboring neurons) to be mostly driven by external biases. The thalamic projections could then play the role of bias where they seed the state of the network. From both anatomical studies and electrophysiological investigations \citep{Groh_2013}, we know that thalamus is at a prime position to modify the signal based on the cognitive processing that is happening in the cortex \citep{Schmitt_2017,Bolkan_2017}.This thalamic-driven regulation entails ``binding in time'' since MD-like thalamus modifies its output cortex at a given time and is itself influenced by what is perceived by the cortex in time prior. But how can the ``binding in time'' avoid locking-in the thalamic function to a set of inputs at a given time? How can thalamus constantly be both ahead of cortex and yet keep track of the past information? The secret may be embedded in the non-recurrent intrinsic structure of thalamus, the recurrent structure of the higher cortical areas, , and the phase-sensitive detection that biases and binds the locally recurrent activity in cortex, with large-scale feedback loops.

\begin{widetext}

\begin{figure}[h!]
\begin{center}
\includegraphics[width=0.6\columnwidth]{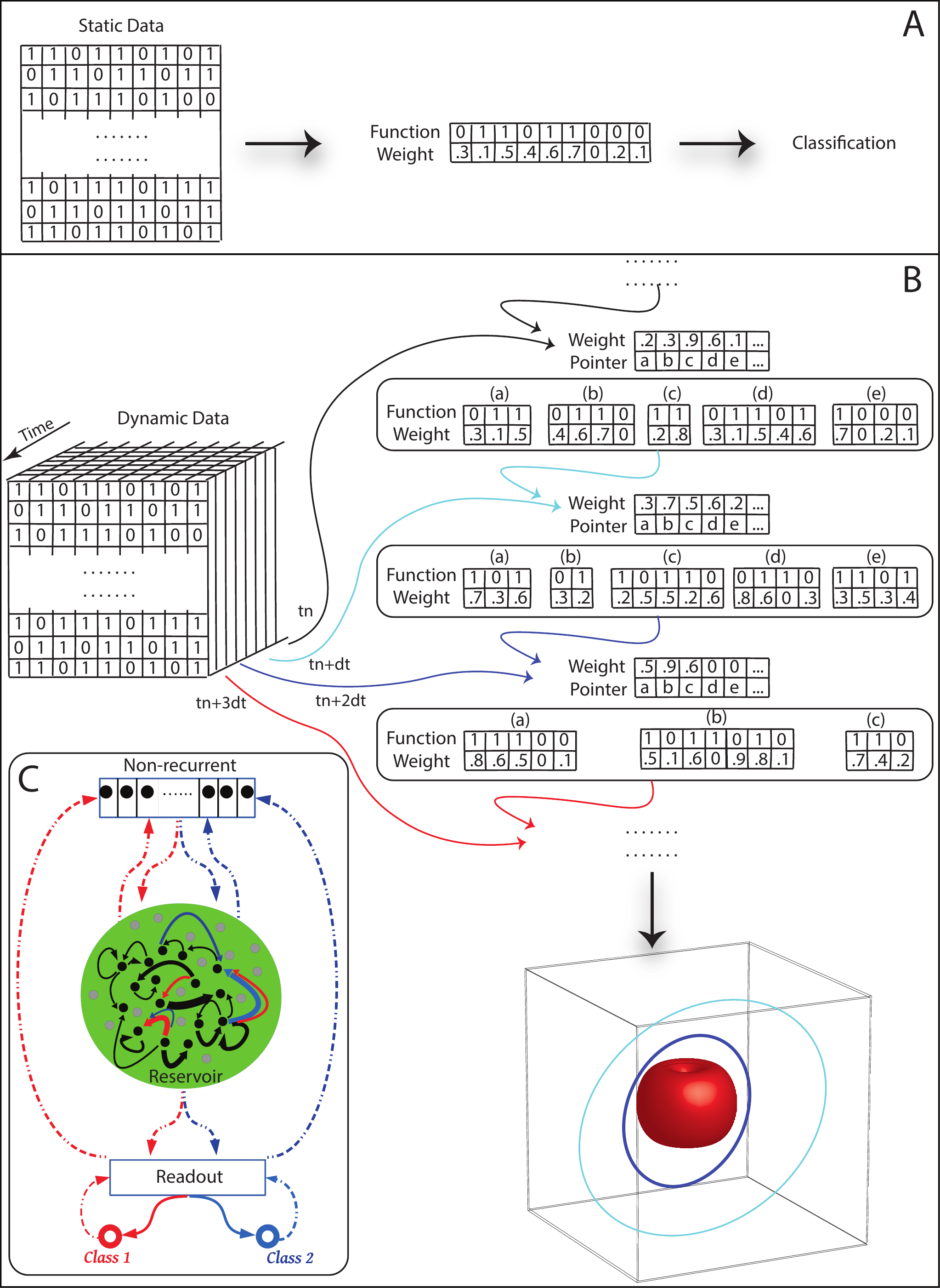}
\caption{{\textbf{Schematic representation of thalamic cognitive contextual
computation.} \emph{(A)}: In the case of static data, a set of
function/weight modules can yield good classification. Function
represents a polynomial (since any system that is known to be a
polynomial Hamiltonians of low order can be accurately approximated by
neural networks \citep{Lin_2017}) and the weights exemplify the
connection matrix of an artificial neural network encapsulating this
polynomial. Stacking multiple of such module can increase the accuracy
of polynomial approximation (such as in the case of CNN). \emph{(B)}:
Thalamo-cortical computation for contextual processing of dynamic data.
Each dataframe is processed by a weight/pointer module (thalamus 
MD-like structure) which like a blackboard is writable by different sets
of neuronal assemblies in cortex. Thalamic pointers assign the
assemblies; modules' weights adjust the influence of each assembly in
further computational step [\emph{inset C} shows 
a non-recurrent thalamic nuclei (MD-like) modulating the weights
in the PFC (reservoir and readout). Here, depending on the context (blue or red), the interactions between MD and Reservoir, between Reservoir and Readout, and between Readout and MD could pursue one of the two possible outcomes. Specifically, MD changes the weights in the Reservoir to differentially set assemblies that produce two different attractor states, each leading to one of the two possible network outputs]. In \emph{(B, C)}, each operation of the thalamic module is itself
influenced, not only by the current frame (\(t\)), but also
by the computation carried by cortex module on the prior frame
(\(t-1\)). Cortical module is composed of multiple assemblies
where each operate similar to the function/weight module of the static
case. These assemblies are locally recurrent and each cell may be
recruited to a different assembly during each operation. This mechanism
could explain why prefrontal cells show mixed selectivity in their
responses to stimuli (as reported in \citep{Rigotti2013,Fusi2016}). Through this
recursive interaction between thalamus and cortex, cognition emerges not
as just a pattern matching computation, but through contextual
computation of dynamic data (bottom right schematic drawing).
{\label{computation}}%
}}
\end{center}
\end{figure}
\end{widetext}

To expand the idea further, let's revisit some core attributes of cognitive processing. Based on the observations of behavior, higher cognition requires ``efficient computation'', ``time delay feedback'', the capacity to ``retain information'' and ``contextual'' computational properties. Such computational cognitive process surpass the computational capacity of simple RNN-like networks. The essential required properties of a complex cognitive system of such kind are: 1) input should be nonlinearly mapped onto the high-dimensional state, while different inputs map onto different states, 2) slightly different states should map onto identical targets, 3) only recent past should influence the state and network is essentially unaware of remote past, 4) a phase-locked loop should decode information that is already encoded in time and 5) the combination of 1-4, should optimize sensory processing based on the context. The first three attributes of such system have close relevance to constrains and computational properties of higher cortical areas (prefrontal). The same three are also the main features of reservoir computing, namely ``separation property'', ``approximation property'' and ``fading memory'' \citep{Jaeger2001,Jaeger2007,Maass2002,Maass2003}. Interestingly, and RC system can ``non-linearly'' map a lower dimensional system to a high-dimensional space facilitating classification of the elements of the low-dimensional space. The last two properties match the structure and computational constraints of non-relay thalamic system as a contextual modulator that is phasically changing the input to the RC system. In fact, in an RC model of prefrontal cortex, addition of a phase neuron significantly improved the networks performance in complex cognitive tasks. The phase neuron improves the performance by generating input driven attractor dynamics that best matched the input \citep{Enel_2016}. This advantageous phase-based bias effect is not limited to the simulation or physiological RC-like neural circuitry. In a recent study, electronic implementation and numerical studies of a limited RC system of a single nonlinear node with delayed feedback has shown efficient information processing  \citep{Appeltant2011}. Such reservoir's transient dynamical response follows delay-dynamical systems, and only a limited set of parameters are required to set the rich dynamical properties of delay systems \citep*{Ikeda_1987}. This system was able to effectively process time-dependent signals. 

The phase neuron \citep{Enel_2016} and delayed dynamical RC \citep{Appeltant2011} both show properties that resemble the structure-function of MD-like thalamus as discussed here. Specifically, the phasic recruitment of cortical neurons is invoked due to the combination of cortical influence on non-relay thalamic neurons through direct and indirect corticothalamic projections (see reticular nucleus inhibitory influence on thalamic neurons, Fig.\ref{Fig xxx}). In a given cycle of computation (Fig.~{\ref{computation}}), cortical feedback leads to the release of GABA (via reticular nucleus) in MD \citep{Kim2011}. It has been shown that the increased GABA alters the opening of T-type Ca2+ channels \citep{Crunelli1991}, which in the case of MD, results in enhanced MD-PFC interaction; yielding mutual drive of the corticothalamic and thalamocortical activity together \citep{Kim2011}. Through this calcium-based low-threshold spiking, the gradual synchronization of MD and PFC ensues \citep{Jones2002, Kim2011}. As a result, MD units and PFC show strong phase-locked synchrony \citep{Parnaudeau2013}. This gradual phase-locking mechanism forms the basis of the temporal dynamics that can nonlinearly (through consecutive cycles of computation) change the cortical activity as a result of novel stimuli or unexpected contextual changes as observed experimentally \citep{Schmitt_2017,Bolkan_2017}. In fact, abnormal activity of T-type Ca2+ channels in MD, leads to hypersynchrony in PFC neurons and frontal lobe-specific seizures\citep{Kim2011}. The interactions between non-relay thalamus and cortex, collectively, is neither feedforward, nor locally recurrent, but it has a mixture of non-recurrent phase encoder that keeps copies of the past processing and modulates the sensory input relay and its next step processing (Fig.~{\ref{computation}}). The distinctive short-term dynamics are well matched with the divergent structure-function relationship of sensory and non-relay MD-like thalamic nuclei. These features further emphasize that the perceptual and cognitive processing can not be solely cortico-centric operations.\\


\section*{Biological constrains and the role of thalamus in computational optimization}

Computation and optimization are two sides of the same coin. But how
does the brain optimize the computations that would match its required
objective, i.e. cognitive processing? There is a current trend of
thinking that brain optimizes some arbitrary functions, with the hopes
that the future discovery of these unknown functions may guide us to
establish a link between brain's operations and deep
learning \citep{Marblestone2016}. This line of approach to optimizational
(and computational) operations of the brain has few flaws. First, it
avoids specifying what function the brain is supposed to optimize (and
as a result it remains vague). Second, it refrains from addressing
certain limitations that brain has to cope with due to biological
constrains. First of these limitations is the importance of using just
enough resources to solve the current perceptual problem. Second is the
necessity to come up with a solution just in (the needed) time. The
importance of ``just-enough'' and ``just-in-time'' computation in
cortical computation should not be overlooked \citep*{Douglas2007}. If the
first condition is not met, the organism can not sustain continued
activity since the metabolic demand surpasses the dedicated energetic
expenditure and the animal can not survive. In fact, the communication
in neural networks are highly constrained by number of factors,
specifically the energetic demands of the network
operations \citep*{Laughlin2003}. From estimates of the cost of cortical
computation \cite{Lennie2003}, we know that the high cost of spiking
forces the brain to rely on sparse communication and using only a small
fraction of the available neurons \citep{Shoham_2006,Baddeley_1997}. While,
theoretically, cortex can dedicate a large number of neurons (and very
high dynamical space) to solve any cognitive task, metabolic demand of
such high-energetic neural activity renders such mechanism highly
inefficient. As a result, the ``law of diminishing returns'' dictates
that increased energetic cost causing excessive pooling of active
neurons to an assembly would be penalized  \citep*{Niven_2008}. The
penalization for unnecessary high-energetic neural activity, in itself,
should be driven by the nature of computation rather than being
formulated as a fixed arbitrary threshold imposed by an external
observer. On the other hand, a system can resort to low-cost computation
at any given time but dedicate long enough time to solve the task on
hand. Naturally, such system would not be very relevant to the
biological systems since time is of essence. If an animal dedicates a
long instance of its computational capacity to solve a problem, the
environment has changed before it reaches a solution and the solution
becomes obsolete. A deer would never have an advantage for its brain to
have fully analyzed the visual scene instead of spotting the approaching
wolf and shifting resources to the most-needed task, i.e. escape. As a
result, many of the optimization techniques and concepts that may be
relevant to artificial neural networks are irrelevant to embodied
computational cognition of the brain. The optimization that the brain
requires is not aiming for the best possible performance, but rather
needs to reach a good mixture of economy and efficiency. 

Not surprisingly, these constrains, i.e. efficiency and economy, are
cornerstones of homeostasis and are observed across many scales in
living systems \citep{Szekely_2013}. The simple ``Integral feedback'' acts
as the mainstay of control feedback in such homeostatic systems (such as
E Coli heat-shock or DNA repair after exposure to gamma
radiation) \citep{EL_SAMAD_2002,El_Samad_2005,Krishna_2007,Dekel_2005}. Change in input leads to change in the
output and the proportional change in the controller aiming to reset the
output to the desired regime. 
When the integral feedback is disrupted, the system can no longer reach proper homeostasis and either efficiency or economy (or even both) will be sub-optimal \citep{EL_SAMAD_2002,El_Samad_2005,Szekely_2013}. Many different etiologies could be behind the integral feedback disruption, but the outcome is loss of robust response in uncertain environments. The presence of feedforward and feedback loops provide the means for robust and fast operation in processing fluctuating incoming inputs. This feedback regulation and operational robustness has an energetic and computational cost for the system. Although for simple systems it is feasible to associate the exact cost of an operation to the overall computational cost of the system, scaling the metabolic cost of feedback regulations to large networks remains a challenge since it will involve multiple feedback loops, nonlinear dynamics and numerous uncertain parameters \citep{Csete2002}. Specifically in the case of a single neurons, branching architecture, non-uniform ion channel distributions and conduction states of action potentials affect the rate of energy consumption \citep{Ju2016}. However, this electrochemical energy of single neuron operation does not linearly scale to the spent energy at networks level \citep{Wang2009,Wang2014}. The total energy function of neural populations will depend not only on the energy function of single neurons and their coupling in a given neural population, but also on the flow of information between different populations \citep{Wang2014,Wang2016}. When a large pool of neurons is recruited to form multiple assemblies to perform a certain computation, it is the interactions between the assemblies that will define the collective behavior of the discrete components. Since the coupled processes show additive entropy productions \citep{Demirel2011}, the total energetic optimality of the desired function would depend on the feedback loops between the assemblies and how these feedbacks control the intrinsic energy expenditure of a given assembly. These attributes are inline with the general principle of modular composition of biological systems \citep{Hartwell1999, Dehghani2017,Dehghani2018}. From the dynamical systems' perspective, to understand the operational principles (here, of large assembly of neurons), we do not need to the strip down the assembly to its individual component level (here, individual neurons) \citep{Dehghani2018}. As a result the optimal control at the functional scale of modules where the interaction between the system's modules take place \citep{Csete2002,Dehghani2018}.

The constrains that we discussed above, directly translate to the computational operations of thalamocortical system as we discussed. Instead of just trying to deal with one
fitness function at a time (where the minima of the landscape would be
deemed as ``the'' optima), the brain has to perform a multi-objective
optimization, finding solutions to both metabolic cost (economy) and
just-in-time (efficiency) computation. Thus we can infer that a unique
solution does not exist for such a problem. Rather, any optimization for
computational efficiency will cost us economy and any optimization for
economy will cost us efficiency. In such case, a multi-objective
optimization pareto frontier is desirable. Pareto frontier of
information/cost will be the set of solutions where any other point in
the space is objectively worse for both of the
objectives \citep{Kung1975,Godfrey2006,Szekely_2013}. As a result, the optimization mechanism
should push the system to this frontier. The iterative dynamical interaction between thalamus and cortex seems to provide an elegant solution for this problem (We discuss this in more details below).

In addition to these theoretical rationals, we also wish to point to some observations that support the emergent optimization in the thalamo-cortical system. For example, metabolic studies have shown that following thalamic injuries, a misbalance in cortical metabolism ensues \citep{Baron1986, Baron1992,Larson1998,Levasseur1992}. Moreover, in healthy humans (and not in mood disorder patients), the metabolic rate of thalamus directly relates to the power of cortical oscillations \citep{Lindgren1999}. The misbalance in cortical metabolism have been observed in variety of nuclei damages, but are specially pronounced in mediodorsal,  centre median or pulvinar injuries \citep{Baron1986, Baron1992}. In addition, low-frequency and high-frequency stimulation of MD can induce long-term depression/potentiation or in mPFC (medial PFC); however, the exact sign and magnitude of the differential modulation of thalamo-prefrontal functions under low and high input drive depends on the lack or presence of Muscarinic and Nicotinic modulation \citep{Bueno2012}. Just as thalamic injury/modulation can change the cortical activity and metabolism, cortical injuries (due to stroke for example) can cause an attenuation of the excitatory feedback to thalamus and lead to thalamo-cortical dysrhythmia \citep{vanWijngaarden2016}. Regardless of where the initial injury has occurred, the disrupted thalamo-cortical interaction is conjoined with a misbalance in metabolism. The resultant out of balance activity leads to cognitive disorders that can happen in form of disrupted information processing due to cortical hypersynchrony as a results of excessive thalamic spiking \citep{Kim2011} or faulty modulation of sensory signals and loss of the normal correlation between glucose metabolism in the thalamus and PFC \citep{Byne2001, Katz1996}. The exact celullar/subcellular mechanism that lies beneath the joint fluctuations of firing and metabolic of cortex and thalamus is not very well understood and number of mechanisms may act (not necessarily exclusively). For example, reduced cortical feedback may lead to thalamic hyperpolarization, and the resultant de-inactivation of voltage-gated T-type Ca channels may cause the neurons to switch from tonic spiking to a pathological bursting \citep{vanWijngaarden2016}. Or it could be that the thalamic drive of the inhibitory neurons in the cortex not only directly affect the cortical mode of firing \citep{Fan2017} but also change the glycogenolysis in astrocytes through Vasoactive intestinal peptide (VIP) interneurons \citep{Magistretti2006,Magistretti2015}. Interestingly, and in contrast to the noradrenergic afferent fibers that span horizontally across cortical domains, VIP neurons have a bipolar architecture and therefore their effect is spatially limited \citep{Magistretti2015}, likely correlated to the size of the functional assemblies that are recruited to perform a computational task. Whichever the exact mechanism at the cellular level is, the collective activity of modulatory thalamus and cortex drives the optimization that inherently can not be controlled by the information available at the scale of single neurons and solely in cortex.

\begin{figure}[h!]
\begin{center}
\includegraphics[width=1\columnwidth]{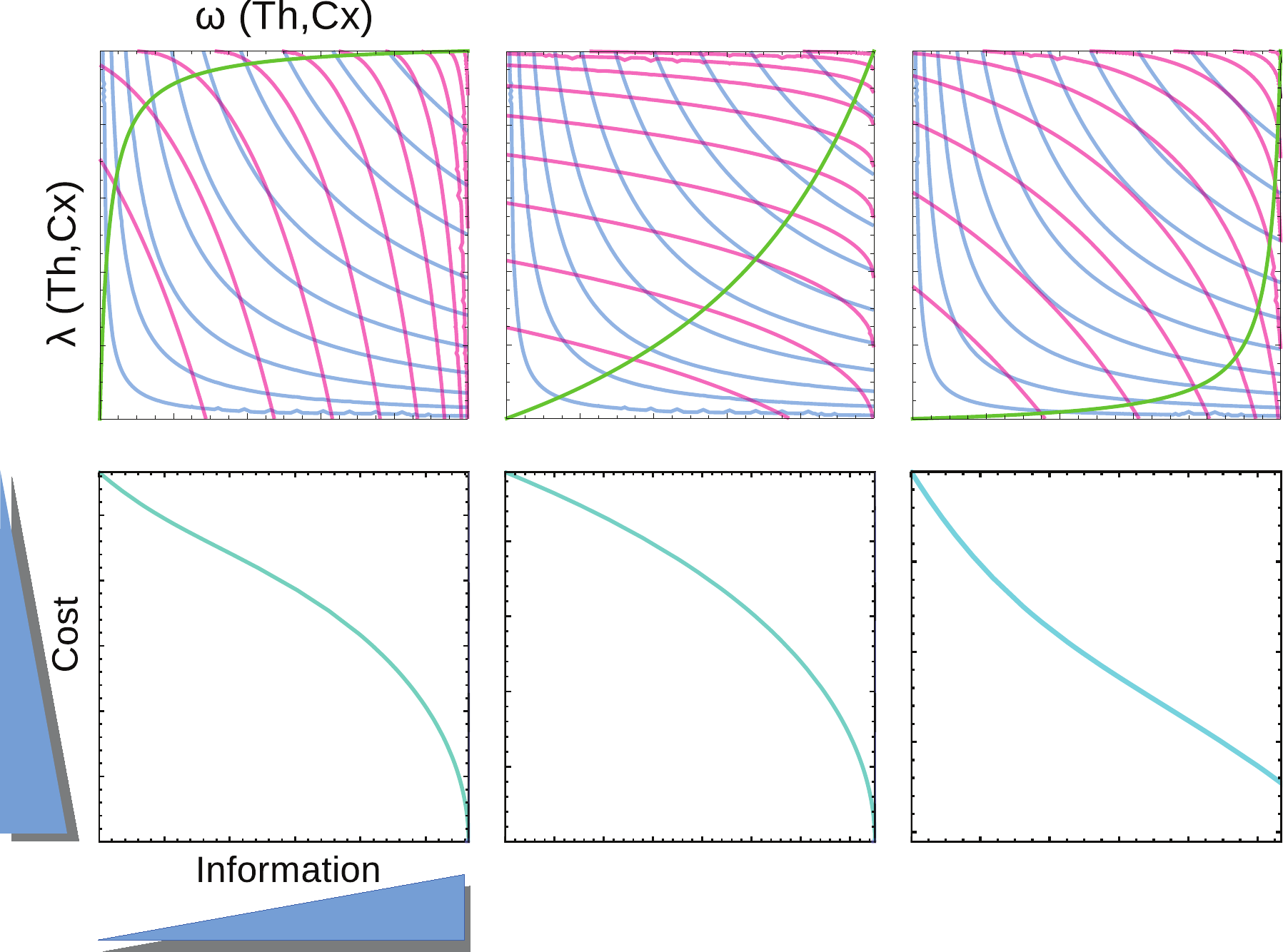}
\caption{{\textbf{Dynamic parameter space of thalamo-cortical joint optimization
of information/cost.} Three different realization of information/cost
interaction as a function of thalamic and cortical activity
(\(\omega,\lambda\)) and the corresponding pareto curves (see
Fig.~{\ref{theory}} for details of this optimization
construct). Pareto curve shows the optimal set of both cost and
information that can be obtained given the biophysical constrains of
neurons and networks connecting them. Every point on the pareto frontier
shows technically efficient levels for a given parameter set
of \(\omega,\lambda\) (see text for more details). All the points inside
the curve are feasible but are not maximally efficient. The slope
(marginal rate of transformation between cost and information) shows how
in order to increase information, cost has to change. The dynamic
nature of interaction between thalamus and cortex enables an emergent
optimization of information/cost depending on the computational problem
on hand and the prior state of the system.
{\label{pareto}}%
}}
\end{center}
\end{figure}

To formalize multiobjective optimization, consider a set of functions, \(\omega\) and \(\lambda\)
of \(f_{Th}\ \) (firing rate of thalamic cell)
and \(f_{Cx}\) (firing rate of cortical cells). Uncertainty (or
its opposite, information) and computational cost (a mixture of time and
metabolic expense) can both be mapped to this functional space
of \(\omega\left(f_{Th},f_{Cx}\right)\) and \(\lambda\left(f_{Th},f_{Cx}\right)\)
(Fig.~{\ref{theory}}A1,2). Let's define computational
cost and information as product and linear sum of cortical and thalamic
activity (\(\ \alpha f_{Th}^n.\beta f_{Cx}^n\),\(\ \theta\frac{d_{fTh}^n}{d_t}+\psi\frac{d_{fCx}^n}{d_t}\); with \(\alpha,\ \beta,\ \theta,\ \psi\)
as coefficients) to reflect the logarithmic nature of information (entropy) and
the fact that biological cost is an accelerating function of the
cost-inducing variables \citep*{Dekel_2005}. The hypothetical space of
cost/information is depicted in Fig.~{\ref{theory}},
where top panels show indifference maps of information (A1) and cost
(A2). The example simulations and parametric plots of the cost and
information functions defined as above are shown in
Fig.~{\ref{pareto}}. In each indifference map, along
each iso-quant curve, the total functional attribute is the same. For
example, anywhere on the \(\eta_1\) curve, the uncertainty (or
information) in our computational engine is the same. However, different
iso-quant curves represent different levels of the functional attribute.
For example, moving outward increases information (reduces uncertainty)
as \(\eta_1<\eta_2<\eta_3<\eta_4\) and thus if computational cost was not a
constrain, the optimal solution would have existed
on \(\eta_4\) or further away
(Fig.{\ref{theory}} A1). In contrast, moving inward
would preserve the cost (\(\zeta_{1\ }<\zeta_2<\zeta_3<\zeta_4\)) if the computational
engine did not have the objective of reducing uncertainty
(Fig.~{\ref{theory}}A2). Since information and cost are
interdependent and both depend on the interaction between thalamus and
cortex, we suggest that information/cost optimization happens through an
iterative interaction between thalamus and cortex (note the blackboard
analogy and contextual modulation discussed above). Since we defined
both information and cost as a set of iso-quant curves in the functional
space of \(\omega\left(f_{Th},f_{Cx}\right)\) and \(\lambda\left(f_{Th},f_{Cx}\right)\), they can be
co-represented in the same space
(Fig.~{\ref{theory}}B). Optimal solutions for
information/cost optimizations are simply the solutions to where the
tangents of the iso-quant curves are equal (see the tangents {[}black
dashed lines{]} and points A, B, C and D, in
Fig.~{\ref{theory}}B). These points create a set of
optimal solutions for the tradeoff between information and cost (green
curve). Mapping of the optimal solutions to the computational efficiency
space \(E\), gives us the pareto efficient curve (cyan
curve, Fig.~{\ref{theory}}C). Anywhere inside the curve
is not pareto efficient (i.e. information gain and computational cost
can change in such a way that, collectively, the system can be in a
better state (on the pareto curve). Points outside of the pareto
efficient curve are not available to the current state of the system due
to the coefficients of \(\omega\) and \(\lambda\). A
change in these coefficients can potentially shape a different
co-representation of information and cost (see
Fig~{\ref{pareto}}, top row for 3 different instances
of \(\omega\) and \(\lambda\) based on
different \(\alpha,\ \beta,\ \theta,\ \psi\) values), and thus a different pareto
efficient curve (see Fig~{\ref{pareto}}, bottom row).
These different possible pareto frontiers can be set based on the prior
state of the system and the complexity of the computational problem on
hand. For example, the modulatory MD-like thalamic triggering of feedforward inhibition of layer I interneurons and layers II/III pyramidals \citep{Cruikshank2012} may tune the cortical activity to a sustained profile of arousal during wakefulness \citep{Harris2011}. Or in contrast to relay thalamus where maximum responsiveness to transient signals (such as sensory stimuli onset/offset) is needed \citep{Rose2005,Bruno2006}, the MD-like modulatory drive may be invoked for tasks where working memory and contextual processing are needed \citep{Parnaudeau2013,Delevich2015}. Nonetheless, the computational efficiency of the system can not be infinitely pushed outward because of the system's intrinsic biophysical constrains (neurons and their wiring). The shaded region in Fig~{\ref{pareto}}, bottom row, shows this non-permissible zone.


In the defined computational efficiency space \(E\),
composed of the two variables information and cost (as the objective
functions, shown in bottom panels of
Fig.~{\ref{theory}} and
Fig.~{\ref{pareto}}), solving a computational problem
is represented by a decrease in uncertainty. However, any change in
uncertainty has an associated cost. First derivative of the pareto
frontier shows ``marginal rate of substitution'' as \(\frac{\Delta_{info}}{\Delta_{cost}}\).
This ratio varies among different points on the pareto efficient curve.
If we take two points on the pareto curve in the computational
efficiency space, such as A and C for example, computational efficiency
of these two points are equal \(E_{A\left(\eta_1,\lambda_4\right)}=\ E_{C\left(\eta_3,\lambda_2\right)}\). The change in
efficiency of point A with respect to information and cost, are the
partial derivatives \( \frac{\partial E_A}{\partial info}\) and \(\frac{\partial E_A}{\partial\cos t}\),
respectively. As a result,\(\frac{\partial E_A}{\partial\inf o}d_{\inf o}+\frac{\partial E_A}{\partial\cos t}d_{\cos t}=0\), meaning that there is
constant efficiency along the pareto curve, the tradeoff between
information and cost is not constant. The optimization in this space is
not based on some fixed built-in algorithm or arbitrary thresholds by an
external observer. Rather, information/cost optimization is the result
of back and forth interaction between thalamus and cortex. Based on the
computational perspective that we have portrayed, thalamus seems to be
poised to operate as an optimizer. Thalamus receives a copy of (sensory)
input while relaying it, and receives an efferent copy from the
processor (cortex), while trying to efficiently bind the information
from past and present and sending it back to cortex. The outcome of such
emergent optimization, is a pareto front in the economy-efficiency
landscape
(Fig.~{\ref{theory}},{\ref{pareto}}).
If the cortex were to be the sole conductor of cognitive processing, the
dynamics of the relay and cortical processing would meander in the
parameter space and not yielding any optimization that can provide a
feasible solution to economic and just-in-time computation. Such system
is doomed to fail, either due to metabolic costs or due to computational
freeze over time ; thus more or less be a useless cognitive engine. In
contrast, with the help of an optimizer that acts as a contextual
modulator, the acceptable parameters will be confined to a manifold
within the parameter space. Such regime would be a sustainable and
favorable domain for cognitive computing. This property shows another 
important facet of a thalamo-cortical computational cognitive system and
the need to move passed the cortico-centric view of cognition. 
An important consequence of this formalization is that it provides us testable hypotheses for objectively evaluating information and cost optimization. By careful simultaneous measurements of thalamic and cortical collective activity, during different states and under different neurotransmitter modulatory effects, one should be able to examine the distinctive interaction of cortex and MD-like versus relay thalamic nuclei. Although we wish to emphasize that while information processing is a fundamentally energy-consuming process \citep{Bennett2003,Parrondo2015} and one can drive theoretical estimates of the energetic cost of the activity of a population of neurons, the exact translation of bit to watts in adaptive information processing systems (such as thalamocortical) can only be verified experimentally \citep{Flack2017}. Without proper and careful measurements, it is impossible to predict how much more reliable the collective computation could get at the expense of energy \citep{Ay2007}. Likewise, the degree to which the energy is traded for accuracy/speed (or their combination) will be a hard challenge for the experimentalists measuring the collective activity \citep{Lan2012}. 
 \\


\section*{Concluding remarks: Reframing Thalamic function above and beyond information relay}
Lately, new evidence about the possible role of thalamus has started to challenged the cortico-centric view of perception/cognition. Anatomical studies and physiological measurements have begun to unravel the importance of the Cortico-Thalamo-Cortical loops in cognitive processes \citep{Basso_2005,Parnaudeau2017}. Under this emerging paradigm, thalamus plays two distinctive roles: a) information relay, b) modulation of cortical function \citep*{Sherman_2013}, where the neocortex does not work in isolation but is largely dependent on thalamus. In contrast to cortical networks which operate as specialized memory devices via their local recurrent excitatory connections, the thalamus is devoid of local connections, and is instead optimized for capturing state information that is distributed across multiple cortical nodes while animals are engaged in context-dependent task switching \citep*{Schmitt_2017}. This allows the thalamus to explicitly represent task context (corresponding to different combinations of cortical states), and through its unique
projection patterns to the cortex, different thalamic inputs modify the effective connections between cortical neurons \citep{Schmitt_2017,Bolkan_2017}.

\begin{widetext}

\begin{figure}[h!]
\begin{center}
\includegraphics[width=0.6\columnwidth]{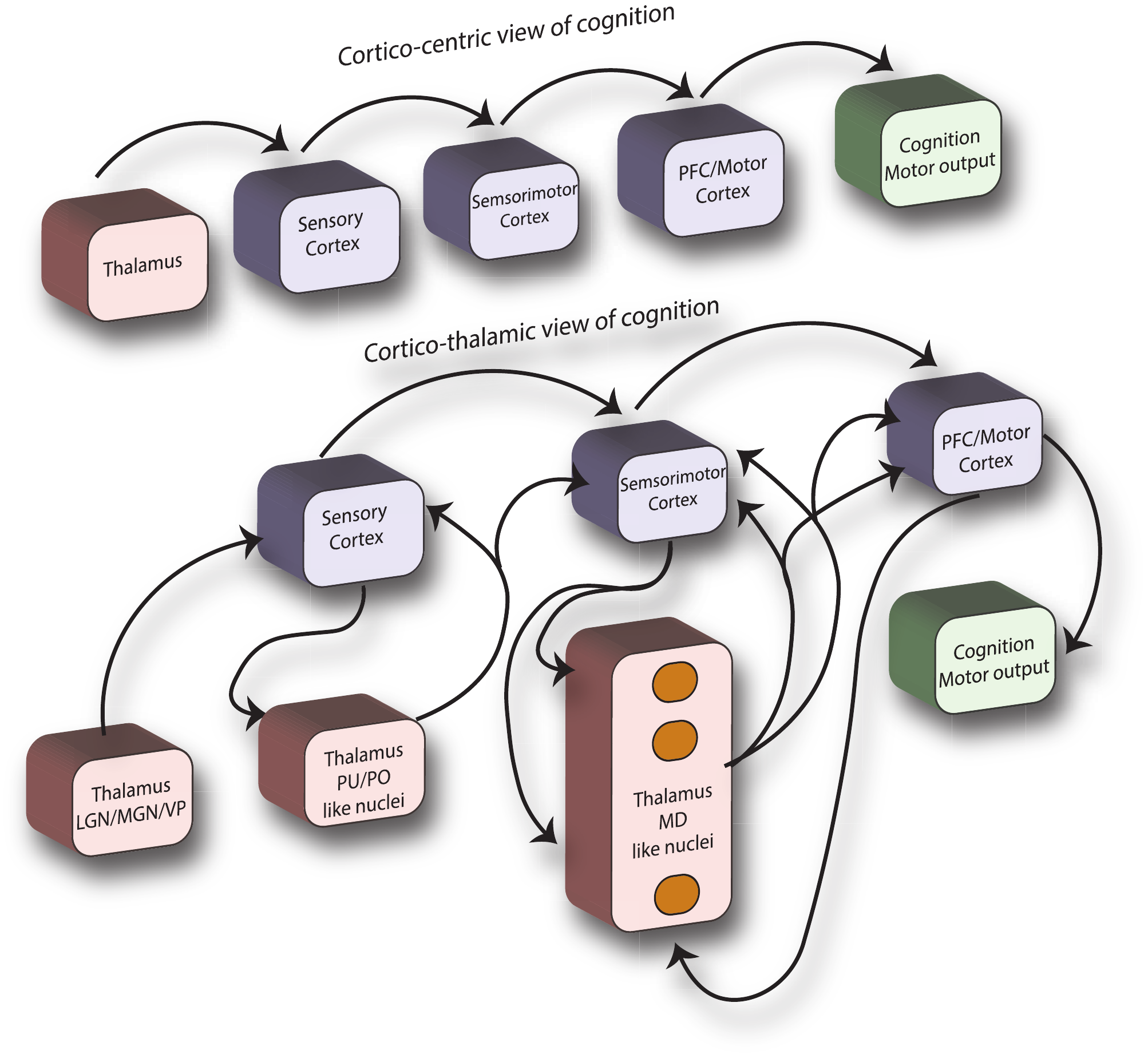}
\caption{{\textbf{The emergent view of thalamic role in cognition.} \emph{(Top)}
In the traditional view, serial processing of information confines the
role of thalamus to only a relay station. \emph{(Bottom)} the view that
is discussed in this manuscript considers thalamus as a key player in
cognition, above and beyond relay to sensory cortices. Through combining
the efferent readout from cortex with sensory afferent, MD-like thalamic
nuclei modulate further activity of the higher cortex. The contextual
modulation enabled by MD is composed of distinctively parallel
operations (individual circles represent the non-recurrent nature of
these processes due to lack of local excitatory connections). Under this
view, and the computational operatives discussed here, the
thalamo-cortical system (and not just cortex) is in charge of contextual
cognitive computing. The computation enabled by Pulvinar/PO like nuclei
is different from LGN and also from MD-like nuclei.
{\label{viewpoint}}%
}}
\end{center}
\end{figure}
\end{widetext}

Here, we started with a brief overview of the architecture of thalamus, the back and forth communication between thalamus and cortex, then we provided the electrophysiological evidence of thalamic modulatory function, and concluded with a computational frame that encapsulates the architectural and functional attributes of the thalamic role in cognition. In such frame, the computational efficiency of the cognitive computing machinery is achieved through iterative interactions between
thalamus and cortex embedded in the hierarchical organization (Fig.~{\ref{experiment}},~{\ref{computation}}). Under this emergent view, thalamus serves not only as relay, but also as a read/write medium for cortical processing , playing a crucial role in contextual modulation of cognition (Fig.~{\ref{viewpoint}}). Such multiscale organization of computational processes is a necessary requirement for design of the intelligent systems \cite{Dehghani2017,Simon1969,Simon1962}. Distributed computing in biological systems in most cases operates without central
control \citep*{Navlakha2014}. This is well reflected in the computational perspective that we discussed here. We suggest that through the continuous contextual modulation of cortical activity, thalamus (along with cortex) plays a significant role in emergent optimization of computational efficiency and computational cost. This phenomenon has a deep relation with phase transitions in complex networks. Different states (phases) of the network are associated with the connectivity of
the computing elements (see thalamic weight/pointer and cortical function/weight modules in Fig.~{\ref{computation}}). Interestingly, intrinsic properties of the complex networks do not define the phase transitions in system. Rather, the interplay of the system with its external environment shapes the landscape where phase transitions occur \citep*{Seoane2015}. This parallel in well-studied physical systems and neuronal networks of thalamo-cortical system show the importance of the interplay between thalamus and cortex in cognitive computation and optimization. The proposed frame for contextual cognitive computation and the emergent information/cost optimization in thalamo-cortical system can guide us in designing novel AI architecture. 

\begin{acknowledgements}
We wish to thank Michael Halassa for helpful discussions.
\end{acknowledgements}

\bibliographystyle{acm}

\begin{thebibliography}{100}

\bibitem[Alcaraz et~al., 2018]{Alcaraz_2018}
Alcaraz, F., Fresno, V., Marchand, A.~R., Kremer, E.~J., Coutureau, E., and
  Wolff, M. (2018).
\newblock {Thalamocortical and corticothalamic pathways differentially
  contribute to goal-directed behaviors in the rat}.
\newblock {\em {eLife}}, 7.

\bibitem[Alel{\'u}-Paz and Gim{\'e}nez-Amaya, 2008]{Alelupaz2008}
Alel{\'u}-Paz, R. and Gim{\'e}nez-Amaya, J.~M. (2008).
\newblock The mediodorsal thalamic nucleus and schizophrenia.
\newblock {\em Journal of psychiatry \& neuroscience: JPN}, 33(6):489.

\bibitem[Appeltant et~al., 2011]{Appeltant2011}
Appeltant, L., Soriano, M.~C., Van~der Sande, G., Danckaert, J., Massar, S.,
  Dambre, j., Schrauwen, B., Mirasso, C.~R., and Fischer, I. (2011).
\newblock {Information processing using a single dynamical node as complex
  system}.
\newblock {\em Nature Communications}, 2:468.

\bibitem[Ay et~al., 2007]{Ay2007}
Ay, N., Flack, J., and Krakauer, D.~C. (2007).
\newblock Robustness and complexity co-constructed in multimodal signalling
  networks.
\newblock {\em Philosophical Transactions of the Royal Society B: Biological
  Sciences}, 362(1479):441--447.

\bibitem[Baddeley et~al., 1997]{Baddeley_1997}
Baddeley, R., Abbott, L.~F., Booth, M. C.~A., Sengpiel, F., Freeman, T.,
  Wakeman, E.~A., and Rolls, E.~T. (1997).
\newblock {Responses of neurons in primary and inferior temporal visual
  cortices to natural scenes}.
\newblock {\em Proceedings of the Royal Society B: Biological Sciences},
  264(1389):1775--1783.

\bibitem[Baron et~al., 1986]{Baron1986}
Baron, J., D'antona, R., Pantano, P., Serdaru, M., Samson, Y., and Bousser, M.
  (1986).
\newblock Effects of thalamic stroke on energy metabolism of the cerebral
  cortex: a positron tomography study in man.
\newblock {\em Brain}, 109(6):1243--1259.

\bibitem[Baron et~al., 1992]{Baron1992}
Baron, J., Levasseur, M., Mazoyer, B., Legault-Demare, F., Mauguiere, F.,
  Pappata, S., Jedynak, P., Derome, P., Cambier, J., and Tran-Dinh, S. (1992).
\newblock Thalamocortical diaschisis: positron emission tomography in humans.
\newblock {\em Journal of Neurology, Neurosurgery \& Psychiatry},
  55(10):935--942.

\bibitem[Basso et~al., 2005]{Basso_2005}
Basso, M.~A., Uhlrich, D., and Bickford, M.~E. (2005).
\newblock {Cortical Function: A View from the Thalamus}.
\newblock {\em Neuron}, 45(4):485--488.

\bibitem[Baxter, 2013]{Baxter2013}
Baxter, M. (2013).
\newblock Mediodorsal thalamus and cognition in non-human primates.
\newblock {\em Frontiers in Systems Neuroscience}, 7:38.

\bibitem[Bennett, 2003]{Bennett2003}
Bennett, C.~H. (2003).
\newblock Notes on landauer's principle, reversible computation, and maxwell's
  demon.
\newblock {\em Studies in History and Philosophy of Science Part B: Studies in
  History and Philosophy of Modern Physics}, 34(3):501 -- 510.
\newblock Quantum Information and Computation.

\bibitem[Bickford, 2016]{Bickford_2016}
Bickford, M.~E. (2016).
\newblock {Thalamic Circuit Diversity: Modulation of the Driver/Modulator
  Framework}.
\newblock {\em Frontiers in Neural Circuits}, 9.

\bibitem[Bolkan et~al., 2017]{Bolkan_2017}
Bolkan, S.~S., Stujenske, J.~M., Parnaudeau, S., Spellman, T.~J., Rauffenbart,
  C., Abbas, A.~I., Harris, A.~Z., Gordon, J.~A., and Kellendonk, C. (2017).
\newblock {Thalamic projections sustain prefrontal activity during working
  memory maintenance}.
\newblock {\em Nature Neuroscience}, 20(7):987--996.

\bibitem[Bruno and Sakmann, 2006]{Bruno2006}
Bruno, R.~M. and Sakmann, B. (2006).
\newblock Cortex is driven by weak but synchronously active thalamocortical
  synapses.
\newblock {\em Science}, 312(5780):1622--1627.

\bibitem[Bueno-Junior et~al., 2012]{Bueno2012}
Bueno-Junior, L.~S., Lopes-Aguiar, C., Ruggiero, R.~N., Romcy-Pereira, R.~N.,
  and Leite, J.~P. (2012).
\newblock Muscarinic and nicotinic modulation of thalamo-prefrontal cortex
  synaptic pasticity in vivo.
\newblock {\em PLOS ONE}, 7:1--11.

\bibitem[Byne et~al., 2001]{Byne2001}
Byne, W., Buchsbaum, M.~S., Kemether, E., Hazlett, v., Shinwari, A.,
  Mitropoulou, V., and Siever, L.~J. (2001).
\newblock Magnetic resonance imaging of the thalamic mediodorsal nucleus and
  pulvinar in schizophrenia and schizotypal personality disorder.
\newblock {\em Archives of General Psychiatry}, 58(2):133--140.

\bibitem[Cardin et~al., 2008]{Cardin2008}
Cardin, J.~A., Palmer, L.~A., and Contreras, D. (2008).
\newblock Cellular mechanisms underlying stimulus-dependent gain modulation in
  primary visual cortex neurons in vivo.
\newblock {\em Neuron}, 59:150--160.

\bibitem[Clasca et~al., 2012]{Clasca2012}
Clasca, F., Rubio-Garrido, P., and Jabaudon, D. (2012).
\newblock {Unveiling the diversity of thalamocortical neuron subtypes}.
\newblock {\em Eur J Neurosci}, 35:1524--32.

\bibitem[Collins et~al., 2018]{Collins2018}
Collins, D.~P., Anastasiades, P.~G., Marlin, J.~J., and Carter, A.~G. (2018).
\newblock Reciprocal circuits linking the prefrontal cortex with dorsal and
  ventral thalamic nuclei.
\newblock {\em Neuron}, 98(2):366--379.e4.

\bibitem[Corbetta, 1998]{Corbetta1998}
Corbetta, M. (1998).
\newblock {Frontoparietal cortical networks for directing attention and the eye
  to visual locations: identical, independent, or overlapping neural systems?}
\newblock {\em Proceedings of National Academy of Science}, 95:831--8.

\bibitem[Cruikshank et~al., 2012]{Cruikshank2012}
Cruikshank, S.~J., Ahmed, O.~J., Stevens, T.~R., Patrick, S.~L., Gonzalez,
  A.~N., Elmaleh, M., and Connors, B.~W. (2012).
\newblock Thalamic control of layer 1 circuits in prefrontal cortex.
\newblock {\em Journal of Neuroscience}, 32(49):17813--17823.

\bibitem[Cruikshank et~al., 2007]{Cruikshank2007}
Cruikshank, S.~J., Lewis, T.~J., and Connors, B.~W. (2007).
\newblock Synaptic basis for intense thalamocortical activation of feedforward
  inhibitory cells in neocortex.
\newblock {\em Nature Neuroscience}, 10.

\bibitem[Crunelli and Leresche, 1991]{Crunelli1991}
Crunelli, V. and Leresche, N. (1991).
\newblock A role for gabab receptors in excitation and inhibition of
  thalamocortical cells.
\newblock {\em Trends in Neurosciences}, 14(1):16 -- 21.

\bibitem[Csete and Doyle, 2002]{Csete2002}
Csete, M.~E. and Doyle, J.~C. (2002).
\newblock Reverse engineering of biological complexity.
\newblock {\em Science}, 295(5560):1664--1669.

\bibitem[Dehghani, 2017]{Dehghani2017}
Dehghani, N. (2017).
\newblock {Design of the Artificial: lessons from the biological roots of
  general intelligence}.
\newblock {\em ArXiv}.

\bibitem[Dehghani, 2018]{Dehghani2018}
Dehghani, N. (2018).
\newblock Theoretical principles of multiscale spatiotemporal control of
  neuronal networks: A complex systems perspective.
\newblock {\em Frontiers in Computational Neuroscience}, 12:81.

\bibitem[Dekel and Alon, 2005]{Dekel_2005}
Dekel, E. and Alon, U. (2005).
\newblock {Optimality and evolutionary tuning of the expression level of a
  protein}.
\newblock {\em Nature}, 436(7050):588--592.

\bibitem[Delevich et~al., 2015]{Delevich2015}
Delevich, K., Tucciarone, J., Huang, Z.~J., and Li, B. (2015).
\newblock The mediodorsal thalamus drives feedforward inhibition in the
  anterior cingulate cortex via parvalbumin interneurons.
\newblock {\em Journal of Neuroscience}, 35(14):5743--5753.

\bibitem[Demirel, 2011]{Demirel2011}
Demirel, Y. (2011).
\newblock {\em Energy Coupling}, pages 419--440.
\newblock MIT Press.

\bibitem[Deniau and Chevalier, 1985]{Deniau_1985}
Deniau, J.~M. and Chevalier, G. (1985).
\newblock {Disinhibition as a basic process in the expression of striatal
  functions. {II}. The striato-nigral influence on thalamocortical cells of the
  ventromedial thalamic nucleus}.
\newblock {\em Brain Research}, 334(2):227--233.

\bibitem[Douglas and Martin, 2007]{Douglas2007}
Douglas, R.~J. and Martin, K.~A. (2007).
\newblock {Mapping the Matrix: The Ways of Neocortex}.
\newblock {\em Neuron}, 56(2):226--238.

\bibitem[El-Samad et~al., 2002]{EL_SAMAD_2002}
El-Samad, H.~J., Goff, J.~P., and Khamash, M.~H. (2002).
\newblock {Calcium Homeostasis and Parturient Hypocalcemia: An Integral
  Feedback Perspective}.
\newblock {\em Journal of Theoretical Biology}, 214(1):17--29.

\bibitem[El-Samad et~al., 2005]{El_Samad_2005}
El-Samad, H.~J., Kurata, H., Doyle, J.~C., Gross, C.~A., and Khamash, M.~H.
  (2005).
\newblock {Surviving heat shock: Control strategies for robustness and
  performance}.
\newblock {\em Proceedings of the National Academy of Sciences},
  102(8):2736--2741.

\bibitem[Enel et~al., 2016]{Enel_2016}
Enel, P., Procyk, E., Quilodran, R., and Dominey, P.~F. (2016).
\newblock {Reservoir Computing Properties of Neural Dynamics in Prefrontal
  Cortex}.
\newblock {\em {PLOS} Computational Biology}, 12(6):e1004967.

\bibitem[Fan et~al., 2017]{Fan2017}
Fan, D., Duan, L., Wang, Q., and Luan, G. (2017).
\newblock Combined effects of feedforward inhibition and excitation in
  thalamocortical circuit on the transitions of epileptic seizures.
\newblock {\em Frontiers in Computational Neuroscience}, 11:59.

\bibitem[Felleman and Essen, 1991]{Felleman_1991}
Felleman, D.~J. and Essen, D. C.~V. (1991).
\newblock {Distributed Hierarchical Processing in the Primate Cerebral Cortex}.
\newblock {\em Cerebral Cortex}, 1(1):1--47.

\bibitem[FitzGibbon et~al., 2015]{FitzGibbon2015}
FitzGibbon, T., Eriköz, B., Grünert, U., and Martin, P.~R. (2015).
\newblock {Analysis of the lateral geniculate nucleus in dichromatic and
  trichromatic marmosets}.
\newblock {\em Journal of Comparative Neurology}, 523(13):1948--1966.

\bibitem[Flack, 2017]{Flack2017}
Flack, J. (2017).
\newblock {\em Life's Information Hierarchy}, page 283–302.
\newblock Cambridge University Press.

\bibitem[Funahashi and Nakamura, 1993]{Funahashi1993}
Funahashi, K. and Nakamura, Y. (1993).
\newblock {Approximation of dynamical systems by continuous time recurrent
  neural networks}.
\newblock {\em Neural Networks}, 6(6):801--806.

\bibitem[Fusi et~al., 2016]{Fusi2016}
Fusi, S., Miller, E.~K., and Rigotti, M. (2016).
\newblock {Why neurons mix: high dimensionality for higher cognition}.
\newblock {\em Current Opinion in Neurobiology}, 37:66--74.

\bibitem[Gabernet et~al., 2005]{Gabernet2005}
Gabernet, L., Jadhav, S.~P., Feldman, D.~E., Carandini, M., and Scanziani, M.
  (2005).
\newblock Somatosensory integration controlled by dynamic thalamocortical
  feed-forward inhibition.
\newblock {\em Neuron}, 48(2):315--327.

\bibitem[Giguere and Goldman-Rakic, 1988]{Giguere1988}
Giguere, M. and Goldman-Rakic, P.~S. (1988).
\newblock Mediodorsal nucleus: Areal, laminar, and tangential distribution of
  afferents and efferents in the frontal lobe of rhesus monkeys.
\newblock {\em Journal of Comparative Neurology}, 277(2):195--213.

\bibitem[Godfrey et~al., 2006]{Godfrey2006}
Godfrey, P., Shipley, R., and Gryz, J. (2006).
\newblock {Algorithms and analyses for maximal vector computation}.
\newblock {\em The {VLDB} Journal}, 16(1):5--28.

\bibitem[Goldberg et~al., 2013]{Goldberg_2013}
Goldberg, J.~H., Farries, M.~A., and Fee, M.~S. (2013).
\newblock {Basal ganglia output to the thalamus: still a paradox}.
\newblock {\em Trends in Neurosciences}, 36(12):695--705.

\bibitem[Goldman-Rakic and Porrino, 1985]{Goldman1985}
Goldman-Rakic, P.~S. and Porrino, L.~J. (1985).
\newblock The primate mediodorsal (md) nucleus and its projection to the
  frontal lobe.
\newblock {\em Journal of Comparative Neurology}, 242(4):535--560.

\bibitem[Grant et~al., 2012]{Grant_2012}
Grant, E., Hoerder-Suabedissen, A., and Moln{\'{a}}r, Z. (2012).
\newblock {Development of the Corticothalamic Projections}.
\newblock {\em Frontiers in Neuroscience}, 6.

\bibitem[Groh et~al., 2014]{Groh_2013}
Groh, A., Bokor, H., Mease, R.~A., Plattner, V.~M., Hangya, B., Stroh, A.,
  Deschenes, M., and Acs{\'{a}}dy, L. (2014).
\newblock {Convergence of Cortical and Sensory Driver Inputs on Single
  Thalamocortical Cells}.
\newblock {\em Cerebral Cortex}, 24(12):3167--3179.

\bibitem[Halassa and Acs{\'{a}}dy, 2016]{Halassa_2016}
Halassa, M.~M. and Acs{\'{a}}dy, L. (2016).
\newblock {Thalamic Inhibition: Diverse Sources Diverse Scales}.
\newblock {\em Trends in Neurosciences}, 39(10):680--693.

\bibitem[Halassa and Kastner, 2017]{Halassa2017}
Halassa, M.~M. and Kastner, S. (2017).
\newblock {Thalamic functions in distributed cognitive control}.
\newblock {\em Nature Neuroscience}, 20(12):1669--1679.

\bibitem[Harris and Thiele, 2011]{Harris2011}
Harris, K.~D. and Thiele, A. (2011).
\newblock Cortical state and attention.
\newblock {\em Nature Reviews Neuroscience}, 12:509 EP --.

\bibitem[Harth et~al., 1987]{Harth1987}
Harth, E.~M., Unnikrishnan, K.~P., and Pandya, A.~S. (1987).
\newblock {The inversion of sensory processing by feedback pathways: a model of
  visual cognitive functions}.
\newblock {\em Science}, 237(4811):184--187.

\bibitem[Hartwell et~al., 1999]{Hartwell1999}
Hartwell, L.~H., Hopfield, J.~J., Leibler, S., and Murray, A.~W. (1999).
\newblock From molecular to modular cell biology.
\newblock {\em Nature}, 402:C47 EP --.

\bibitem[Heeger, 2017]{Heeger2017}
Heeger, D.~J. (2017).
\newblock {Theory of cortical function}.
\newblock {\em Proceedings of the National Academy of Sciences},
  114(8):1773--1782.

\bibitem[Hubel and Wiesel, 1959]{Hubel1959}
Hubel, D.~H. and Wiesel, T.~N. (1959).
\newblock {Receptive fields of single neurones in the cat{\textquotesingle}s
  striate cortex}.
\newblock {\em The Journal of Physiology}, 148(3):574--591.

\bibitem[Hubel and Wiesel, 1962]{Hubel1962}
Hubel, D.~H. and Wiesel, T.~N. (1962).
\newblock {Receptive fields binocular interaction and functional architecture
  in the cat{\textquotesingle}s visual cortex}.
\newblock {\em The Journal of Physiology}, 160(1):106--154.

\bibitem[Ikeda and Matsumoto, 1987]{Ikeda_1987}
Ikeda, K. and Matsumoto, K. (1987).
\newblock {High-dimensional chaotic behavior in systems with time-delayed
  feedback}.
\newblock {\em Physica D: Nonlinear Phenomena}, 29(1-2):223--235.

\bibitem[Jaeger, 2001]{Jaeger2001}
Jaeger, H. (2001).
\newblock {The echo state approach to analysing and training recurrent neural
  networks}.
\newblock Technical report.

\bibitem[Jaeger, 2007]{Jaeger2007}
Jaeger, H. (2007).
\newblock {Echo state network}.
\newblock {\em Scholarpedia}, 2(9):2330.

\bibitem[Jazayeri and Shadlen, 2015]{Jazayeri_2015}
Jazayeri, M. and Shadlen, M.~N. (2015).
\newblock {A Neural Mechanism for Sensing and Reproducing a Time Interval}.
\newblock {\em Current Biology}, 25(20):2599--2609.

\bibitem[Jones, 1981]{Jones1981}
Jones, E.~G. (1981).
\newblock {Functional subdivision and synaptic organization of the mammalian
  thalamus}.
\newblock {\em Int Rev Physiol}, 25:173--245.

\bibitem[Jones, 1985]{Jones1985}
Jones, E.~G. (1985).
\newblock {Principles of Thalamic Organization}.
\newblock In {\em The Thalamus}, pages 85--149. Springer {US}.

\bibitem[Jones, 1998]{Jones1998}
Jones, E.~G. (1998).
\newblock {Viewpoint: the core and matrix of thalamic organization}.
\newblock {\em Neuroscience}, 85(2):331--345.

\bibitem[Jones, 2002]{Jones2002}
Jones, E.~G. (2002).
\newblock Thalamic circuitry and thalamocortical synchrony.
\newblock {\em Philosophical Transactions of the Royal Society of London B:
  Biological Sciences}, 357(1428):1659--1673.

\bibitem[Ju et~al., 2016]{Ju2016}
Ju, H., Hines, M.~L., and Yu, Y. (2016).
\newblock Cable energy function of cortical axons.
\newblock {\em Scientific Reports}, 6:29686 EP --.

\bibitem[Kakei et~al., 2001]{Kakei2001}
Kakei, S., Na, J., and Shinoda, Y. (2001).
\newblock {Thalamic terminal morphology and distribution of single
  corticothalamic axons originating from layers 5 and 6 of the cat motor
  cortex}.
\newblock {\em The Journal of Comparative Neurology}, 437(2):170--185.

\bibitem[Katz et~al., 1996]{Katz1996}
Katz, M., Buchsbaum, M.~S., Siegel~Jr, B.~V., Wu, J., Haier, R.~J., and
  Bunney~Jr, W.~E. (1996).
\newblock Correlational patterns of cerebral glucose metabolism in
  never-medicated schizophrenics.
\newblock {\em Neuropsychobiology}, 33(1):1--11.

\bibitem[Khubieh et~al., 2016]{Khubieh2016}
Khubieh, A., Ratté, S., Lankarany, M., and Prescott, S.~A. (2016).
\newblock Regulation of cortical dynamic range by background synaptic noise and
  feedforward inhibition.
\newblock {\em Cerebral Cortex}, 26(8):3357--3369.

\bibitem[Kim et~al., 1995]{Kim1995}
Kim, H.~G., Beierlein, M., and Connors, B.~W. (1995).
\newblock Inhibitory control of excitable dendrites in neocortex.
\newblock {\em Journal of Neurophysiology}, 74(4):1810--1814.

\bibitem[Kim et~al., 2011]{Kim2011}
Kim, J., Woo, J., Park, Y.-G., Chae, S., Jo, S., Choi, J.~W., Jun, H.~Y., Yeom,
  Y.~I., Park, S.~H., Kim, K.~H., Shin, H.-S., and Kim, D. (2011).
\newblock Thalamic t-type ca2+ channels mediate frontal lobe dysfunctions
  caused by a hypoxia-like damage in the prefrontal cortex.
\newblock {\em Journal of Neuroscience}, 31(11):4063--4073.

\bibitem[Komura et~al., 2013]{Komura_2013}
Komura, Y., Nikkuni, A., Hirashima, N., Uetake, T., and Miyamoto, A. (2013).
\newblock {Responses of pulvinar neurons reflect a subject{\textquotesingle}s
  confidence in visual categorization}.
\newblock {\em Nature Neuroscience}, 16(6):749--755.

\bibitem[Krettek and Price, 1977]{Krettek1977}
Krettek, J.~E. and Price, J.~L. (1977).
\newblock The cortical projections of the mediodorsal nucleus and adjacent
  thalamic nuclei in the rat.
\newblock {\em Journal of Comparative Neurology}, 171(2):157--191.

\bibitem[Krishna et~al., 2007]{Krishna_2007}
Krishna, S., Maslov, S., and Sneppen, K. (2007).
\newblock {{UV}-Induced Mutagenesis in Escherichia coli {SOS} Response: A
  Quantitative Model}.
\newblock {\em {PLoS} Computational Biology}, 3(3):e41.

\bibitem[Kung et~al., 1975]{Kung1975}
Kung, H.-T., Luccio, F., and Preparata, F.~P. (1975).
\newblock {On Finding the Maxima of a Set of Vectors}.
\newblock {\em Journal of the {ACM}}, 22(4):469--476.

\bibitem[Kuramoto et~al., 2017]{Kuramoto2016}
Kuramoto, E., Pan, S., Furuta, T., Tanaka, Y.~R., Iwai, H., Yamanaka, A., Ohno,
  S., Kaneko, T., Goto, T., and Hioki, H. (2017).
\newblock {Individual mediodorsal thalamic neurons project to multiple areas of
  the rat prefrontal cortex: A single neuron-tracing study using virus
  vectors}.
\newblock {\em Journal of Comparative Neurology}, 525(1):166--185.

\bibitem[Kuroda et~al., 1998]{Kuroda1998}
Kuroda, M., Yokofujita, J., and Murakami, K. (1998).
\newblock An ultrastructural study of the neural circuit between the prefrontal
  cortex and the mediodorsal nucleus of the thalamus.
\newblock {\em Progress in Neurobiology}, 54(4):417 -- 458.

\bibitem[Kuroda et~al., 2004]{Kuroda2004}
Kuroda, M., Yokofujita, J., Oda, S., and Price, J.~L. (2004).
\newblock Synaptic relationships between axon terminals from the mediodorsal
  thalamic nucleus and $\gamma$-aminobutyric acidergic cortical cells in the
  prelimbic cortex of the rat.
\newblock {\em Journal of Comparative Neurology}, 477(2):220--234.

\bibitem[Lan et~al., 2012]{Lan2012}
Lan, G., Sartori, P., Neumann, S., Sourjik, V., and Tu, Y. (2012).
\newblock The energy--speed--accuracy trade-off in sensory adaptation.
\newblock {\em Nature Physics}, 8:422 EP --.

\bibitem[Larson et~al., 1998]{Larson1998}
Larson, C.~L., Davidson, R.~J., Abercrombie, H.~C., Ward, R.~T., Schaefer,
  S.~M., Jackson, D.~C., Holden, J.~E., and Perlman, S.~B. (1998).
\newblock Relations between pet-derived measures of thalamic glucose metabolism
  and eeg alpha power.
\newblock {\em Psychophysiology}, 35(2):162--169.

\bibitem[Laughlin and Sejnowski, 2003]{Laughlin2003}
Laughlin, S.~B. and Sejnowski, T.~J. (2003).
\newblock {Communication in neuronal networks}.
\newblock {\em Science}, 301:1870--4.

\bibitem[Lennie, 2003]{Lennie2003}
Lennie, P. (2003).
\newblock {The Cost of Cortical Computation}.
\newblock {\em Current Biology}, 13(6):493--497.

\bibitem[Levasseur et~al., 1992]{Levasseur1992}
Levasseur, M., Baron, J., Sette, G., Legault-Demare, F., Pappata, S.,
  Mauguiere, F., Benoit, N., Dinh, S.~T., Degos, J., Laplane, D., et~al.
  (1992).
\newblock Brain energy metabolism in bilateral paramedian thalamic infarcts: a
  positron emission tomography study.
\newblock {\em Brain}, 115(3):795--807.

\bibitem[Lewis et~al., 2012]{Lewis2012}
Lewis, D.~A., Curley, A.~A., Glausier, J.~R., and Volk, D.~W. (2012).
\newblock Cortical parvalbumin interneurons and cognitive dysfunction in
  schizophrenia.
\newblock {\em Trends in Neurosciences}, 35(1):57--67.

\bibitem[Lin et~al., 2017]{Lin_2017}
Lin, H.~W., Tegmark, M., and Rolnick, D. (2017).
\newblock {Why Does Deep and Cheap Learning Work So Well?}
\newblock {\em Journal of Statistical Physics}, 168(6):1223--1247.

\bibitem[Lindgren et~al., 1999]{Lindgren1999}
Lindgren, K.~A., Larson, C.~L., Schaefer, S.~M., Abercrombie, H.~C., Ward,
  R.~T., Oakes, T.~R., Holden, J.~E., Perlman, S.~B., Benca, R.~M., and
  Davidson, R.~J. (1999).
\newblock Thalamic metabolic rate predicts eeg alpha power in healthy control
  subjects but not in depressed patients.
\newblock {\em Biological Psychiatry}, 45(8):943--952.

\bibitem[Ma and Jazayeri, 2014]{Ma_2014}
Ma, W.~J. and Jazayeri, M. (2014).
\newblock {Neural Coding of Uncertainty and Probability}.
\newblock {\em Annual Review of Neuroscience}, 37(1):205--220.

\bibitem[Maass et~al., 2003]{Maass2003}
Maass, W., Natschlaeger, T., and Markram, H. (2003).
\newblock {Computational Models for Generic Cortical Microcircuits}.
\newblock In {\em Computational Neuroscience}. Chapman and Hall/{CRC}.

\bibitem[Maass et~al., 2002]{Maass2002}
Maass, W., Natschläger, T., and Markram, H. (2002).
\newblock {Real-Time Computing Without Stable States: A New Framework for
  Neural Computation Based on Perturbations}.
\newblock {\em Neural Computation}, 14(11):2531--2560.

\bibitem[Magistretti, 2006]{Magistretti2006}
Magistretti, P.~J. (2006).
\newblock Neuron{\textendash}glia metabolic coupling and plasticity.
\newblock {\em Journal of Experimental Biology}, 209(12):2304--2311.

\bibitem[Magistretti and Allaman, 2015]{Magistretti2015}
Magistretti, P.~J. and Allaman, I. (2015).
\newblock A cellular perspective on brain energy metabolism and functional
  imaging.
\newblock {\em Neuron}, 86(4):883 -- 901.

\bibitem[Mante et~al., 2013]{Mante2013}
Mante, V., Sussillo, D., Shenoy, K.~V., and Newsome, W.~T. (2013).
\newblock {Context-dependent computation by recurrent dynamics in prefrontal
  cortex}.
\newblock {\em Nature}, 503(7474):78--84.

\bibitem[Marblestone et~al., 2016]{Marblestone2016}
Marblestone, A.~H., Wayne, G., and Kording, K.~P. (2016).
\newblock {Toward an Integration of Deep Learning and Neuroscience}.
\newblock {\em Frontiers in Computational Neuroscience}, 10.

\bibitem[Marenco et~al., 2012]{Marenco_2011}
Marenco, S., Stein, J.~L., Savostyanova, A.~A., Sambataro, F., Tan, H.-Y.,
  Goldman, A.~L., Verchinski, B.~A., Barnett, A.~S., Dickinson, D., Apud,
  J.~A., Callicott, J.~H., Meyer-Lindenberg, A., and Weinberger, D.~R. (2012).
\newblock {Investigation of Anatomical Thalamo-Cortical Connectivity and {fMRI}
  Activation in Schizophrenia}.
\newblock {\em Neuropsychopharmacology}, 37(2):499--507.

\bibitem[Mathers, 1972]{Mathers_1972}
Mathers, L.~H. (1972).
\newblock {The synaptic organization of the cortical projection to the pulvinar
  of the squirrel monkey}.
\newblock {\em The Journal of Comparative Neurology}, 146(1):43--59.

\bibitem[Mesulam, 1990]{Mesulam1990}
Mesulam, M.~M. (1990).
\newblock {Large-scale neurocognitive networks and distributed processing for
  attention, language, and memory}.
\newblock {\em Ann Neurol}, 28:597--613.

\bibitem[Mitchell, 2015]{Mitchell2015}
Mitchell, A.~S. (2015).
\newblock {The mediodorsal thalamus as a higher order thalamic relay nucleus
  important for learning and decision-making.}
\newblock {\em Neurosci Biobehav Rev}, 54:76--88.

\bibitem[Mitelman et~al., 2005]{Mitelman_2005}
Mitelman, S.~A., Byne, W., Kemether, E.~M., Hazlett, E.~A., and Buchsbaum,
  M.~S. (2005).
\newblock {Metabolic Disconnection Between the Mediodorsal Nucleus of the
  Thalamus and Cortical Brodmann's Areas of the Left Hemisphere in
  Schizophrenia}.
\newblock {\em American Journal of Psychiatry}, 162(9):1733--1735.

\bibitem[Mumford, 1991]{Mumford1991}
Mumford, D. (1991).
\newblock {On the computational architecture of the neocortex. I: The role of
  the thalamo-cortical loop}.
\newblock {\em Biological Cybernetics}, 65(2):135--145.

\bibitem[Mumford, 1992]{Mumford1992}
Mumford, D. (1992).
\newblock {On the computational architecture of the neocortex. II The role of
  cortico-cortical loops}.
\newblock {\em Biological Cybernetics}, 66(3):241--251.

\bibitem[Nair et~al., 2013]{Nair_2013}
Nair, A., Treiber, J.~M., Shukla, D.~K., Shih, P., and Müller, R.-A. (2013).
\newblock {Impaired thalamocortical connectivity in autism spectrum disorder: a
  study of functional and anatomical connectivity}.
\newblock {\em Brain}, 136(6):1942--1955.

\bibitem[Nakajima and Halassa, 2017]{Nakajima2017}
Nakajima, M. and Halassa, M.~M. (2017).
\newblock {Thalamic control of functional cortical connectivity}.
\newblock {\em Current Opinion in Neurobiology}, 44:127--131.

\bibitem[Navlakha and Bar-Joseph, 2014]{Navlakha2014}
Navlakha, S. and Bar-Joseph, Z. (2014).
\newblock {Distributed information processing in biological and computational
  systems}.
\newblock {\em Communications of the {ACM}}, 58(1):94--102.

\bibitem[Newell, 1962]{Newell1962}
Newell, A. (1962).
\newblock {Some problems of basic organization in problem solving programs}.

\bibitem[Nii, 1986]{Nii1986}
Nii, P. (1986).
\newblock {The Blackboard Model of Problem Solving and the Evolution of
  Blackboard Architectures}.
\newblock {\em The AI Magazine}, 7(2):38--53.

\bibitem[Niven and Laughlin, 2008]{Niven_2008}
Niven, J.~E. and Laughlin, S.~B. (2008).
\newblock {Energy limitation as a selective pressure on the evolution of
  sensory systems}.
\newblock {\em Journal of Experimental Biology}, 211(11):1792--1804.

\bibitem[Ohno et~al., 2012]{Ohno2012}
Ohno, S., Kuramoto, E., Furuta, T., Hioki, H., Tanaka, Y., Fujiyama, F.,
  Sonomura, T., Uemura, M., Sugiyama, K., and Kaneko, T. (2012).
\newblock {A morphological analysis of thalamocortical axon fibers of rat
  posterior thalamic nuclei: a single neuron tracing study with viral vectors.}
\newblock {\em Cereb Cortex}, 22:2840--57.

\bibitem[Parnaudeau et~al., 2017]{Parnaudeau_2017}
Parnaudeau, S., Bolkan, S.~S., and Kellendonk, C. (2017).
\newblock {The Mediodorsal Thalamus: An Essential Partner of the Prefrontal
  Cortex for Cognition}.
\newblock {\em Biological Psychiatry}.

\bibitem[Parnaudeau et~al., 2013]{Parnaudeau2013}
Parnaudeau, S., O'Neill, P.-K., Bolkan, S.~S., Ward, R.~D., Abbas, A.~I., Roth,
  B.~L., Balsam, P.~D., Gordon, J.~A., and Kellendonk, C. (2013).
\newblock Inhibition of mediodorsal thalamus disrupts thalamofrontal
  connectivity and cognition.
\newblock {\em Neuron}, 77(6):1151--1162.

\bibitem[Parnaudeau et~al., 2015]{Parnaudeau_2015}
Parnaudeau, S., Taylor, K., Bolkan, S.~S., Ward, R.~D., Balsam, P.~D., and
  Kellendonk, C. (2015).
\newblock {Mediodorsal Thalamus Hypofunction Impairs Flexible Goal-Directed
  Behavior}.
\newblock {\em Biological Psychiatry}, 77(5):445--453.

\bibitem[Parrondo et~al., 2015]{Parrondo2015}
Parrondo, J. M.~R., Horowitz, J.~M., and Sagawa, T. (2015).
\newblock Thermodynamics of information.
\newblock {\em Nature Physics}, 11:131 EP --.

\bibitem[Partlow et~al., 1977]{Partlow_1977}
Partlow, G.~D., Colonnier, M., and Szabo, J. (1977).
\newblock {Thalamic projections of the superior colliculus in the rhesus
  monkey,Macaca mulatta. A light and electron microscopic study}.
\newblock {\em The Journal of Comparative Neurology}, 171(3):285--317.

\bibitem[Pelzer et~al., 2017]{Pelzer2017}
Pelzer, P., Horstmann, H., and Kuner, T. (2017).
\newblock Ultrastructural and functional properties of a giant synapse driving
  the piriform cortex to mediodorsal thalamus projection.
\newblock {\em Frontiers in Synaptic Neuroscience}, 9:3.

\bibitem[Pinault, 2004]{Pinault_2004}
Pinault, D. (2004).
\newblock {The thalamic reticular nucleus: structure function and concept}.
\newblock {\em Brain Research Reviews}, 46(1):1--31.

\bibitem[Popken et~al., 2000]{Popken2000}
Popken, G.~J., Bunney, W.~E., Potkin, S.~G., and Jones, E.~G. (2000).
\newblock Subnucleus-specific loss of neurons in medial thalamus of
  schizophrenics.
\newblock {\em Proceedings of the National Academy of Sciences},
  97(16):9276--9280.

\bibitem[Prescott and De~Koninck, 2003]{Prescott2003}
Prescott, S.~A. and De~Koninck, Y. (2003).
\newblock Gain control of firing rate by shunting inhibition: Roles of synaptic
  noise and dendritic saturation.
\newblock {\em Proceedings of the National Academy of Sciences},
  100(4):2076--2081.

\bibitem[Purushothaman et~al., 2012]{Purushothaman2012}
Purushothaman, G., Marion, R., Li, K., and Casagrande, V. (2012).
\newblock {Gating and control of primary visual cortex by pulvinar.}
\newblock {\em Nat Neurosci}, 15:905--12.

\bibitem[Raczkowski and Fitzpatrick, 1990]{Raczkowski1990}
Raczkowski, D. and Fitzpatrick, D. (1990).
\newblock {Terminal arbors of individual physiologically identified
  geniculocortical axons in the tree shrew{\textquotesingle}s striate cortex}.
\newblock {\em The Journal of Comparative Neurology}, 302(3):500--514.

\bibitem[Rigotti et~al., 2013]{Rigotti2013}
Rigotti, M., Barak, O., Warden, M.~R., Wang, X.-J., Daw, N.~D., Miller, E.~K.,
  and Fusi, S. (2013).
\newblock {The importance of mixed selectivity in complex cognitive tasks}.
\newblock {\em Nature}, 497(7451):585--590.

\bibitem[Rigotti et~al., 2010]{Rigotti2010}
Rigotti, M., Rubin, D. B.~D., Wang, X.-J., and Fusi, S. (2010).
\newblock {Internal representation of task rules by recurrent dynamics: the
  importance of the diversity of neural responses}.
\newblock {\em Frontiers in Computational Neuroscience}, 4.

\bibitem[Rikhye et~al., 2018]{Rikhye2018}
Rikhye, R.~V., Wimmer, R.~D., and Halassa, M.~M. (2018).
\newblock {Towards an integrative theory of thalamic function}.
\newblock {\em Nature Neuroscience (In press)}, xxx(xxx):xxx--xxx.

\bibitem[Rose and Metherate, 2005]{Rose2005}
Rose, H.~J. and Metherate, R. (2005).
\newblock Auditory thalamocortical transmission is reliable and temporally
  precise.
\newblock {\em Journal of Neurophysiology}, 94(3):2019--2030.

\bibitem[Rotaru et~al., 2005]{Rotaru2005}
Rotaru, D.~C., Barrionuevo, G., and Sesack, S.~R. (2005).
\newblock Mediodorsal thalamic afferents to layer iii of the rat prefrontal
  cortex: Synaptic relationships to subclasses of interneurons.
\newblock {\em Journal of Comparative Neurology}, 490(3):220--238.

\bibitem[Rouiller and Welker, 2000]{Rouiller_2000}
Rouiller, E.~M. and Welker, E. (2000).
\newblock {A comparative analysis of the morphology of corticothalamic
  projections in mammals}.
\newblock {\em Brain Research Bulletin}, 53(6):727--741.

\bibitem[Rovo et~al., 2012]{Rovo2012}
Rovo, Z., Ulbert, I., and Acsády, L. (2012).
\newblock {Drivers of the primate thalamus}.
\newblock {\em J Neurosci}, 32:17894--908.

\bibitem[Saalmann, 2014]{Saalmann2014}
Saalmann, Y.~B. (2014).
\newblock Intralaminar and medial thalamic influence on cortical synchrony,
  information transmission and cognition.
\newblock {\em Frontiers in Systems Neuroscience}, 8:83.

\bibitem[Saalmann and Kastner, 2011]{Saalmann2011}
Saalmann, Y.~B. and Kastner, S. (2011).
\newblock Cognitive and perceptual functions of the visual thalamus.
\newblock {\em Neuron}, 71(2):209 -- 223.

\bibitem[Saalmann and Kastner, 2015]{Saalmann2015}
Saalmann, Y.~B. and Kastner, S. (2015).
\newblock The cognitive thalamus.
\newblock {\em Frontiers in Systems Neuroscience}, 9:39.

\bibitem[Schmitt et~al., 2017]{Schmitt_2017}
Schmitt, L.~I., Wimmer, R.~D., Nakajima, M., Happ, M., Mofakham, S., and
  Halassa, M.~M. (2017).
\newblock {Thalamic amplification of cortical connectivity sustains attentional
  control}.
\newblock {\em Nature}, 545(7653):219--223.

\bibitem[Schwartz et~al., 1991]{Schwartz1991}
Schwartz, M.~L., Dekker, J.~J., and Goldman-Rakic, P.~S. (1991).
\newblock Dual mode of corticothalamic synaptic termination in the mediodorsal
  nucleus of the rhesus monkey.
\newblock {\em Journal of Comparative Neurology}, 309(3):289--304.

\bibitem[Scott et~al., 2017]{Scott2017}
Scott, B.~B., Constantinople, C., Akrami, A., Hanks, T.~D., Brody, C.~D., and
  Tank, D.~W. (2017).
\newblock {Fronto-parietal Cortical Circuits Encode Accumulated Evidence with a
  Diversity of Timescales}.
\newblock {\em Neuron}, 95:385--398.e5.

\bibitem[Selfridge, 1959]{Selfridge1959}
Selfridge, O.~G. (1959).
\newblock {Pandemonium: a paradigm for learning in }.
\newblock In Blake, D.~V. and Uttley, A.~M., editors, {\em Proceedings of the
  Symposium on Mechanisation of Thought Processes}, pages 513--526, London.
  National Physical Laboratory.

\bibitem[Seoane and Sol{\'{e}}, 2015]{Seoane2015}
Seoane, L.~F. and Sol{\'{e}}, R. (2015).
\newblock {Phase transitions in Pareto optimal complex networks}.
\newblock {\em Physical Review E}, 92(3).

\bibitem[Sherman, 2016]{Sherman2016}
Sherman, S.~M. (2016).
\newblock {Thalamus plays a central role in ongoing cortical functioning}.
\newblock {\em Nature Neuroscience}, 16(4):533--541.

\bibitem[Sherman and Guillery, 2002]{Sherman_2002}
Sherman, S.~M. and Guillery, R.~W. (2002).
\newblock {The role of the thalamus in the flow of information to the cortex}.
\newblock {\em Philosophical Transactions of the Royal Society B: Biological
  Sciences}, 357(1428):1695--1708.

\bibitem[Sherman and Guillery, 2013]{Sherman_2013}
Sherman, S.~M. and Guillery, R.~W. (2013).
\newblock {\em {Functional Connections of Cortical Areas}}.
\newblock The {MIT} Press.

\bibitem[Shoham et~al., 2006]{Shoham_2006}
Shoham, S., O'Connor, D.~H., and Segev, R. (2006).
\newblock {How silent is the brain: is there a {\textquotedblleft}dark
  matter{\textquotedblright} problem in neuroscience?}
\newblock {\em Journal of Comparative Physiology A}, 192(8):777--784.

\bibitem[Simon, 1962]{Simon1962}
Simon, H.~A. (1962).
\newblock {The Architecture of Complexity}.
\newblock {\em Proceedings of the American Philosophical Society},
  106(6):467--482.

\bibitem[Simon, 1969]{Simon1969}
Simon, H.~A. (1969).
\newblock {\em {The Sciences of the Artificial}}, chapter The Architecture of
  Complexity.
\newblock MIT Press.

\bibitem[Sipper, 1999]{Sipper1999}
Sipper, M. (1999).
\newblock The emergence of cellular computing.
\newblock {\em Computer}, 32(7):18--26.

\bibitem[Sol{\'e} and Macia, 2013]{Sole2013}
Sol{\'e}, R.~V. and Macia, J. (2013).
\newblock Expanding the landscape of biological computation with synthetic
  multicellular consortia.
\newblock {\em Natural Computing}, 12(4):485--497.

\bibitem[Spacek and Lieberman, 1974]{Spacek1974}
Spacek, J. and Lieberman, A. (1974).
\newblock Ultrastructure and three-dimensional organization of synaptic
  glomeruli in rat somatosensory thalamus.
\newblock {\em Journal of anatomy}, 117(Pt 3):487.

\bibitem[Steriade and Deschenes, 1984]{Steriade1984b}
Steriade, M. and Deschenes, M. (1984).
\newblock The thalamus as a neuronal oscillator.
\newblock {\em Brain Research Reviews}, 8(1):1--63.

\bibitem[Steriade et~al., 1986]{Steriade1986}
Steriade, M., Domich, L., and Oakson, G. (1986).
\newblock {Reticularis thalami neurons revisited: activity changes during
  shifts in states of vigilance}.
\newblock {\em J Neurosci}, 6:68--81.

\bibitem[Steriade and Llin{\'{a}}s, 1988]{Steriade_1988}
Steriade, M. and Llin{\'{a}}s, R.~R. (1988).
\newblock {The functional states of the thalamus and the associated neuronal
  interplay}.
\newblock {\em Physiological Reviews}, 68(3):649--742.

\bibitem[Steriade and Pare, 2007]{Steriade2007}
Steriade, M. and Pare, D. (2007).
\newblock {Morphology and electroresponsive properties of thalamic neurons}.
\newblock In {\em Gating in Cerebral Networks}, pages 1--26. Cambridge
  University Press.

\bibitem[Steriade et~al., 1984]{Steriade1984a}
Steriade, M., Parent, A., and Hada, J. (1984).
\newblock {Thalamic projections of nucleus reticularis thalami of cat: A study
  using retrograde transport of horseradish peroxidase and fluorescent
  tracers}.
\newblock {\em The Journal of Comparative Neurology}, 229(4):531--547.

\bibitem[Szekely et~al., 2013]{Szekely_2013}
Szekely, P., Sheftel, H., Mayo, A., and Alon, U. (2013).
\newblock {Evolutionary Tradeoffs between Economy and Effectiveness in
  Biological Homeostasis Systems}.
\newblock {\em {PLoS} Computational Biology}, 9(8):e1003163.

\bibitem[Uhlhaas and Singer, 2010]{Uhlhaas2010}
Uhlhaas, P.~J. and Singer, W. (2010).
\newblock Abnormal neural oscillations and synchrony in schizophrenia.
\newblock {\em Nature Reviews Neuroscience}, 11:100 EP --.

\bibitem[Urbain and Desch{\^{e}}nes, 2007]{Urbain_2007}
Urbain, N. and Desch{\^{e}}nes, M. (2007).
\newblock {Motor Cortex Gates Vibrissal Responses in a Thalamocortical
  Projection Pathway}.
\newblock {\em Neuron}, 56(4):714--725.

\bibitem[van Wijngaarden et~al., 2016]{vanWijngaarden2016}
van Wijngaarden, J. B.~G., Zucca, R., Finnigan, S., and Verschure, P. F. M.~J.
  (2016).
\newblock The impact of cortical lesions on thalamo-cortical network dynamics
  after acute ischaemic stroke: A combined experimental and theoretical study.
\newblock {\em PLOS Computational Biology}, 12(8):1--16.

\bibitem[Viaene et~al., 2011a]{Viaene2011a}
Viaene, A.~N., Petrof, I., and Sherman, S.~M. (2011a).
\newblock Properties of the thalamic projection from the posterior medial
  nucleus to primary and secondary somatosensory cortices in the mouse.
\newblock {\em Proceedings of the National Academy of Sciences},
  108(44):18156--18161.

\bibitem[Viaene et~al., 2011b]{Viaene2011b}
Viaene, A.~N., Petrof, I., and Sherman, S.~M. (2011b).
\newblock Synaptic properties of thalamic input to the subgranular layers of
  primary somatosensory and auditory cortices in the mouse.
\newblock {\em Journal of Neuroscience}, 31(36):12738--12747.

\bibitem[Wang et~al., 2018]{Wang_2017}
Wang, J., Narain, D., Hosseini, E.~A., and Jazayeri, M. (2018).
\newblock {Flexible timing by temporal scaling of cortical responses}.
\newblock {\em Nature Neuroscience}.

\bibitem[Wang and Wang, 2014]{Wang2014}
Wang, R. and Wang, Z. (2014).
\newblock Energy distribution property and energy coding of a structural neural
  network.
\newblock {\em Frontiers in Computational Neuroscience}, 8:14.

\bibitem[Wang et~al., 2009]{Wang2009}
Wang, R., Zhang, Z., and Chen, G. (2009).
\newblock Energy coding and energy functions for local activities of the brain.
\newblock {\em Neurocomputing}, 73(1):139 -- 150.

\bibitem[Wang and Zhu, 2016]{Wang2016}
Wang, R. and Zhu, Y. (2016).
\newblock Can the activities of the large scale cortical network be expressed
  by neural energy? a brief review.
\newblock {\em Cognitive neurodynamics}, 10(1):1--5.

\bibitem[Woodward et~al., 2017]{Woodward_2017}
Woodward, N.~D., Giraldo-Chica, M., Rogers, B., and Cascio, C.~J. (2017).
\newblock {Thalamocortical Dysconnectivity in Autism Spectrum Disorder: An
  Analysis of the Autism Brain Imaging Data Exchange}.
\newblock {\em Biological Psychiatry: Cognitive Neuroscience and Neuroimaging},
  2(1):76--84.

\bibitem[Yamins and DiCarlo, 2016]{Yamins2016}
Yamins, D. L.~K. and DiCarlo, J.~J. (2016).
\newblock {Using goal-driven deep learning models to understand sensory
  cortex}.
\newblock {\em Nature Neuroscience}, 19(3):356--365.

\bibitem[Yamins et~al., 2014]{Yamins_2014}
Yamins, D. L.~K., Hong, H., Cadieu, C.~F., Solomon, E.~A., Seibert, D., and
  DiCarlo, J.~J. (2014).
\newblock {Performance-optimized hierarchical models predict neural responses
  in higher visual cortex}.
\newblock {\em Proceedings of the National Academy of Sciences},
  111(23):8619--8624.

\bibitem[Zhou et~al., 2016]{Zhou2016}
Zhou, H., Schafer, R.~J., and Desimone, R. (2016).
\newblock Pulvinar-cortex interactions in vision and attention.
\newblock {\em Neuron}, 89(1):209--220.

\bibitem[Zikopoulos and Barbas, 2007]{Zikopoulos2007}
Zikopoulos, B. and Barbas, H. (2007).
\newblock Parallel driving and modulatory pathways link the prefrontal cortex
  and thalamus.
\newblock {\em PLOS ONE}, 2(9):1--19.

\end{thebibliography}

\end{document}